\documentclass[prd,a4paper,nofootinbib, preprintnumbers, amsmath,multicol,12pt]{revtex4}

\usepackage{slashed}
\usepackage{color, verbatim, array}
\usepackage{latexsym}
\usepackage{epsfig}
\usepackage{amsmath}
\usepackage{amssymb}
\usepackage{graphicx}
\usepackage{bm}
\usepackage[normalem]{ulem}

\makeatletter

\newcommand\ee{\end{equation}}
\newcommand\be{\begin{equation}}
\newcommand\eea{\end{eqnarray}}
\newcommand\bea{\begin{eqnarray}}

\newcommand\QQm{\mathcal Q_m}

\def\beq{\begin{equation}}
\def\eeq{\end{equation}}

\begin{document}
\setcounter{page}{0}
\thispagestyle{empty}

\noindent CP3-Origins-2015-5  $^{}$\hfill  DESY 15-027\\
DIAS-2015-5 
$^{}$\hfill UAB-FT-767\\



\begin{center}

{\Large\sc \color{red}  
Electroweak vacuum stability\\[-4.5mm]and\\ finite
quadratic radiative corrections
}
\vspace{0.5cm}

{\large  
Isabella Masina$^{a,b}$, Germano Nardini$^{c}$ and Mariano Quiros$^{d}$  }
\\[.5cm]
{\normalsize \small { \sl
$^{a}$ Dip. di Fisica e Scienze della Terra, Universit\`a di Ferrara and INFN Sez. di Ferrara, \\ \vspace{-.2cm}Via Saragat 1, I-44100 Ferrara, Italy}}\\
{\normalsize \small { \sl 
$^{b}$ CP$^\mathbf 3$-Origins and DIAS, Southern Denmark University, \\ \vspace{-.2cm}Campusvej 55, DK-5230 Odense M, Denmark}}\\
{\normalsize \small { \sl
$^{c}$ Deutsches Elektronen Synchrotron, Notkestrasse 85, D-22603 Hamburg, Germany  
}}\\
{\normalsize \small { \sl
$^{d}$ Instituci\'o Catalana de Recerca i Estudis  
Avan\c{c}ats (ICREA) and\\ \vspace{-.2cm}IFAE-UAB 08193 Bellaterra, Barcelona, Spain
}}\\

\end{center}
\vspace{.3cm}

\setcounter{page}{1}

\begin{abstract}\baselineskip=15pt
  \begin{center} {\bf Abstract} \end{center} If the Standard Model
  (SM) is an effective theory, as currently believed, it is valid up
  to some energy scale $\Lambda$ to which the Higgs vacuum expectation
  value is sensitive throughout radiative quadratic terms. The latter
  ones destabilize the electroweak vacuum and generate the SM
  hierarchy problem. For a given perturbative Ultraviolet (UV)
  completion, the SM cutoff can be computed in terms of fundamental
  parameters. If the UV mass spectrum involves several scales the
  cutoff is not unique and each SM sector has its own UV cutoff
  $\Lambda_i$. We have performed this calculation assuming the Minimal
  Supersymmetric Standard Model (MSSM) is the SM UV completion.  As a
  result, from the SM point of view, the quadratic corrections to the
  Higgs mass are equivalent to finite threshold contributions. For the
  measured values of the top quark and Higgs masses, and depending on
  the values of the different cutoffs $\Lambda_i$, these contributions
  can cancel even at renormalization scales as low as multi-TeV,
  unlike the case of a single cutoff where the cancellation only
  occurs at Planckian energies, a result originally obtained by
  Veltman.  From the MSSM point of view, the requirement of stability
  of the electroweak minimum under radiative corrections is
  incorporated into the matching conditions and provides an extra
  constraint on the Focus Point solution to the little hierarchy
  problem in the MSSM. These matching conditions can be employed for
  precise calculations of the Higgs sector in scenarios with heavy
  supersymmetric fields.
  
\end{abstract}
\maketitle 


\baselineskip=16pt

\vspace{.8cm}


\newpage
\section{\sc Introduction}
\label{introduction}

In the Standard Model (SM)~\cite{SM} as an effective field theory with a
physical cutoff $\Lambda$, the Higgs mass parameter is corrected by
(uncalculable) quadratic divergences that can destabilize the
electroweak vacuum. This fact is usually associated with the SM
hierarchy problem~\cite{Gildener:1976ih}.  In the presence of a
perturbative Ultraviolet (UV) completion (beyond the TeV scale) with
heavy fields coupled to the Higgs sector, the quadratic divergences
appear as finite threshold effects, which can therefore be reliably
computed in perturbation theory, after the heavy states are integrated
out at the matching scale between the Low-Energy (LE) 
and the UV High-Energy (HE) effective theories.  So, if the UV completion of the SM
is known and it is perturbative, the hierarchy problem is entirely due to
calculable finite effects and can be fully quantified.  If the UV
completion is non-perturbative, as it happens in the case where the
Higgs is composite, the calculation cannot rely on perturbation
theory, but the presence of a new scale, even if it is dynamically
generated, makes it possible to estimate the size of the threshold
corrections to the Higgs mass~\cite{Tavares:2013dga}.  Here we will
consider the former case where the UV completion is perturbative.

The absence of any departure from the SM predictions in current
experimental data at the LHC is pointing towards the existence of new
physics at least in the multi-TeV region, by which the naturalness
problem is becoming more acute.  This in turn is hinting at less
conventional solutions to the hierarchy problem, as e.g. hypothetical
solutions provided by the theory which breaks supersymmetry at the
(high) scale where supersymmetry breaking is transmitted from the
hidden to the observable sector.  It is therefore important to compute
the large radiative contributions to the hierarchy problem in order
to 
settle the required conditions at the high scale for the stability of the electroweak
vacuum in the effective theory (the SM) below the
matching scale.

In this paper we will consider the SM as the LE effective theory of
the Minimal Supersymmetric Standard Model (MSSM), matching the two
theories at the decoupling scale $\mathcal Q_m$ where the
supersymmetric partners are integrated out.  We assume the MSSM is
valid up to scales of the order of the Planck scale $M_P$, where it
can be understood as e.g.~the flat limit ($M_P\to\infty$) of $N=1$
supergravity ~\cite{Nilles:1983ge}, which should eventually be in turn
UV completed by some more fundamental (superstring) theory.  The hope
is that the fundamental theory could provide the requirements for
solving the SM hierarchy problem, under the form of some HE parameter
relations. For that reason, in this paper we are trying to fix the
required conditions which could lead to stability of the electroweak minimum, but by no means
are we trying to claim any solution to the hierarchy problem, nor even
a precise quantification of the fine-tuning.

Threshold effects when matching the SM with the MSSM have been
extensively studied for dimensionless parameters, as e.g.~the SM
Yukawa and quartic couplings, thus fixing the physical Higgs and fermion
masses~\cite{Draper:2013oza}.  For dimensional parameters, as the
Higgs mass parameter, the thresholds have not been systematically
considered~\footnote{For a previous analysis in the MSSM in the broken
  electroweak symmetric phase, see Ref.~\cite{Baer:2012cf}.}.
However, the hierarchy problem precisely resides in those dimensional
parameters, as solving the equations of minimum providing the
electroweak vacuum expectation value (VEV) of the Higgs field requires a certain amount
of fine-tuning which quantifies the hierarchy problem~\footnote{The
  instability under radiative corrections coming from new heavy
  physics, which is the subject of the present paper, should not be
  confused with the instability driven by the
  Renormalization Group (RG) running of the quartic Higgs coupling
  towards negative values~\cite{Sher:1993mf}.}.

It is thus worth improving our knowledge on the dimensional parameter thresholds. 
To this aim, in this paper we analyze the effects on the SM Higgs mass parameter 
from the decoupling of the heavy MSSM fields. 
This will result in precise relations to be satisfied at the matching scale $\mathcal Q_m$, in order to have the stability of the electroweak minimum. 
Technically, we perform the matching in the one-loop RG-improved approximation, as going beyond one-loop should not add
qualitative complications or dominant contributions. Similarly we perform the matching procedure in the symmetric phase for the MSSM, so our results should be affected by tiny corrections of $\mathcal O(v^2/\mathcal Q_m^2)$.

The outline of the paper goes as follows. In section~\ref{decoupling}
we present some general ideas about the decoupling using the scale
invariance of the effective potential in the one-loop RG-improved
approximation. We show that in the considered approximation the
decoupling scale $\mathcal Q_m$ is arbitrary, although in view of
minimizing (unconsidered) higher loop corrections it is convenient to
take it of the order of magnitude of the masses of the decoupled
fields. Simple toy models to illustrate the general matching procedure
are presented in section~\ref{toy}, where we also stress the role, for
scale invariance, of the anomalous dimensions of scalar fields
(included in the wave-function radiative corrections) which is an
ingredient alien to the effective potential, constructed at zero
external momentum. The case of the MSSM is reviewed in
section~\ref{MSSM} and the detailed matching between SM and MSSM Higgs
mass parameters is performed in section~\ref{radiative}. The threshold
effects induced in the effective theory are computed in
section~\ref{quadratic}. In particular we show that for the MSSM
scenario with degenerate soft breaking masses the finite correction to
the SM Higgs mass parameter precisely reproduces the result obtained
by Veltman (in the context of dimensional regularization in two
dimensions)~\cite{Veltman:1980mj} if the SM cutoff is identified with
the common mass of the degenerate and heavy supersymmetric
partners. Instead, for the more general scenario with non degenerate
heavy masses, the effective theory can be often interpreted as a SM
with different cutoffs for each (quarks, gauge bosons, \dots)
sector. In such a case the finite correction to the SM Higgs mass
parameter consists in a generalized Veltman result with deformations
that can be negligible depending on the hierarchy of the spectrum. In
section~\ref{numerical} we express the HE parameters evaluated at the
decoupling scale $\mathcal Q_m$ in terms of their values at the scale
$\mathcal M$ where the supersymmetry breaking is transmitted to the
observable sector, and we constrain these values to be compatible with
sensible matching conditions. We then quantify the thresholds effects
and their impact on the Higgs sector. We focus on the parameter region
corresponding to the Focus Point (FP) solutions. The regions where the
electroweak stability is not spoiled by the finite corrections in the
matching conditions can be understood as a generalized FP solutions
which include thresholds effects. Our conclusions are presented in
section~\ref{conclusion}, as well as a discussion on the scale, gauge
and renormalization scheme dependence of our results. Finally,
technical details of the calculation of the radiative corrections to
the SM Higgs mass parameter stemming from the different supersymmetric
sectors are presented in section~\ref{appendixA}. A nice check of the
consistency of our calculation is the explicit proof of the one-loop
scale invariance of our results, and this is presented in
section~\ref{appendixB}.

\section{\sc General ideas about the decoupling}
\label{decoupling}
\noindent
Before putting forward the explicit relation of the effective
potential $V$ in the LE and HE regions, let us review some
general ideas about the effective potential~\cite{Ford:1992mv}. The
effective potential improved by the RG depends, on top of the
background value of fields $\phi_i$, on a number of running parameters
$\lambda_I$ (they include dimensionless couplings as well as
dimensionful parameters) and on the renormalization scale $\mathcal
Q$, in such a way that the equation
 \be
\left[\mathcal Q\frac{\partial}{\partial\mathcal Q}+\beta_I\frac{\partial}{\partial\lambda_I}+\gamma_i\phi_i\frac{\partial}{\partial\phi_i}\right]V=0\ 
\label{independence}
 \ee
 is fulfilled. This equation, where $\gamma_i$ are the anomalous dimensions of the fields $\phi_i$, and $\beta_I$ the $\beta$-functions of the parameters $\lambda_I$, highlights the renormalization-scale independence of the effective potential. The general solution of Eq.~\eqref{independence} reads as
 \be
 V=V(\mathcal Q(t),\lambda_I(t),\phi_i(t))
 \label{potencial}
 \ee
 where
 \begin{equation}
 \mathcal Q(t)=\mathcal Q \exp(t)\,,~~ \beta_I(\lambda_I(t))=\frac{d \lambda_I(t)}{dt}\,,~~
 \gamma_i(\lambda_I(t))=\frac{d\log (\phi_i(t)/\phi_i)}{dt}\,,
 \label{defRun}
 \end{equation}
together with the boundary conditions $\lambda_I(0) \equiv \lambda_I$, $\phi_i(0) \equiv \phi_i$.
 
In practice the scale independence of the effective potential (\ref{potencial}) holds up to the level of perturbation theory where the potential is computed. In particular if we make a loop expansion of the operator $V$ as
 \be
 V=\sum_{\ell\geq 0} V^{(\ell)} \ ,
 \label{expansion}
 \ee
 the RG-improved potential $V^{(0)}$ has a very strong scale dependence. This dependence is reduced by considering $V^{(0)}+V^{(1)}$, where $V^{(1)}$ includes the terms that correspond to the field redefinitions $\phi_i(t)\equiv (1+Z_i^{(1)}(t))\phi_i$, and the one-loop RG-improved Coleman-Weinberg contribution $\Delta V^{(1)}$. In fact, in $V^{(0)}+V^{(1)}$ the whole one-loop scale
dependence cancels out and different choices of $Q$ only affect higher order corrections.  Whereas the explicit expression of $Z_i(t)$ depends on the specific model, the contribution $\Delta V^{(1)}$ can be generically written as~\cite{Coleman:1973jx}~\footnote{As customary, in Eq.~\eqref{CW} the matrix $\mathcal M^2$ is the squared mass spectrum in the presence of the background fields $\phi_i$, and the diagonal matrix $\mathcal C$ depends on the renormalization-scheme. Note that in Eq.~\eqref{CW} the $t$-dependence is implicit. Concerning the radiative corrections $Z_i(t)$, we remind that they appear due to the canonical normalization of the kinetic terms and $Z_i(0)$ can contain finite contributions which depend on the renormalization scheme.}
 \be
\Delta V^{(1)}=\frac{1}{64\pi^2}STr \mathcal M^4(\phi_i)\left[ \log\frac{\mathcal M^2(\phi_i)}{\mathcal Q^2}-\mathcal C \right]\ 
 \label{CW}
 \ee
where $STr$ includes the number of degrees of freedom of the different mass eigenstates as well as a negative sign for fermions.
 
 The electroweak-breaking condition, described as the solution to the equations of minimum 
 \be
 \frac{\partial V}{\partial\phi_i} =0\ ,
 \ee
 is also scale independent. Such condition can thus be deduced from
 the potential $V$ at any scale $\mathcal Q$ and any loop order.  However since one is
 only able to compute $\Delta V^{(\ell)}$ and $Z_i^{(\ell)}$ with
 $\ell\leq n$ (i.e.~up to some perturbative order $n$), and the
 minimization condition should eventually be related to electroweak
 observables, in practical cases it is advisable to minimize
 $\sum_{\ell\leq n} V^{(\ell)}$ (with scale dependence at $n+1$ loop
 order) at the electroweak scale $\mathcal Q=\mathcal
 Q_{EW}$. Employing this choice of renormalization scale is subtle
 when also heavy fields are involved, as we describe now.
  
 We consider a HE theory with light and heavy fields. Light fields have electroweak-breaking and/or invariant masses of order (or below) $\mathcal Q_{EW}$, whereas heavy fields have masses $M\gg \mathcal Q_{EW}$. To any fixed order in the loop expansion heavy fields can induce large logarithms in the
 minimization condition evaluated at $\mathcal Q= \mathcal Q_{EW}$~\cite{Bando:1992wy}. For this reason, the minimization should still be performed at $\mathcal Q= \mathcal Q_{EW}$, but in the LE effective theory where the heavy fields have been decoupled.

As we are considering mass-independent renormalization schemes the decoupling of heavy fields has to be performed at some scale $\mathcal Q_m$. In such a case the effective description at $\mathcal Q\ll\mathcal Q_{m}$ is obtained in two steps: {\it i)} Matching at $\mathcal Q=\mathcal Q_m$ of the HE Lagrangian to an effective Lagrangian (which has all light-fields interactions allowed by the HE symmetry), and {\it ii)} Running of the effective couplings from $\mathcal Q_m$ to $\mathcal Q_{EW}$.  The matching of the couplings of the light scalar sector can be obtained by exploiting the LE and HE effective potentials.

 By construction, the HE and LE theories (in the presence of only light-field backgrounds) have the same RG-improved potentials at the decoupling scale, i.e.~$V_{LE}(\mathcal Q_m)=V(\mathcal Q_m)$ (LE quantities carry a ``$LE$'' subscript; for HE quantities the subscript ``$HE$'' is suppressed).  In the ideal case of perfect scale invariance this equivalence is true at any scale, and the choice of $\mathcal Q_m$ at which one matches the two potentials is fully arbitrary. However, in realistic situations where the potentials are calculated at a given loop approximation, $\mathcal Q_m$ has to be set at a value that presumably minimizes the unknown higher order
 corrections coming from heavy fields. This motivates the choice $\mathcal Q_m\sim M$.

 The decoupling procedure involves some technical details when applied to realistic frameworks.  The first issue arises when the HE theory contains several Higgses acquiring VEVs. This complication does not lead to any difficulties in our analysis as we are considering scenarios with the SM as an effective description. In fact we are restricting ourselves to cases where all the multi-Higgs VEVs in the interaction-eigenstate basis can be aligned along a unique VEV in the mass-eigenstate basis (and this direction corresponds to the light Higgs of the SM). Clearly this is possible because we are assuming the mass of the extra-Higgses to be
 $M\gg \mathcal Q_{EW}$.  A second issue arises when there are several heavy fields with masses $M_I$. If they differ by orders of magnitude, there exists no choice of $\mathcal Q_m$ that avoids large logarithms in the matching conditions. In this case the decoupling procedure has to be repeated as many times as the number of hierarchically different heavy mass thresholds. Of course when all $M_I$ are similar, the decoupling can be performed just once with $\mathcal Q_m$ fixed at some intermediate value among $M_I$. The simplified concrete examples of the next sections will better clarify the details of the decoupling procedure.

 \section{\sc Decoupling in some toy models}
\label{toy}
\noindent
In this section we illustrate the previous ideas about decoupling. We first analyze a toy model with only one heavy degree of freedom. Second we consider a case with several scalar heavy particles and light fermions which contribute to the light scalar wave function renormalization. The reader not interested in those technical details can jump straightforwardly to section~\ref{MSSM}.

\subsection{First toy model: scalars}
\noindent
We consider a toy model consisting of a light scalar $\phi$ and a heavy scalar $S$ with a HE Lagrangian
\beq
\mathcal L=-\Omega+\frac{1}{2}(\partial\phi)^2+\frac{1}{2}(\partial S)^2-\frac{1}{2}m^2\phi^2-\frac{1}{4!}\lambda\phi^4-\frac{1}{2}M^2 S^2-\frac{1}{2}h^2\phi^2S^2
\eeq
where for simplicity the quartic coupling of the $S$ field has been
set to zero, although it is not protected by any symmetry. After
decoupling the $S$-field the theory is described by the effective
Lagrangian
\beq
\mathcal L_{LE}=-\Omega_{LE}+\frac{1}{2}(\partial\phi_{LE})^2-\frac{1}{2}m_{LE}^2\phi_{LE}^2-\frac{1}{4!}\lambda_{LE}\phi_{LE}^4~.
\eeq
In both Lagrangians the parameters are running with the scale $\mathcal Q$.

Since $S$ does not acquire a VEV, the electroweak breaking field $\phi$ is aligned to $\phi_{LE}$. The tree-level (RG-improved) matching of the parameters at the scale $\mathcal Q_m$ is then trivial: 
\be
\Omega_{LE}(\mathcal Q_m)=\Omega(\mathcal Q_m),\ \phi_{LE}(\mathcal Q_m)=\phi(\mathcal Q_m) ,\ m_{LE}(\mathcal
Q_m)=m(\mathcal Q_m) ,\ \lambda_{LE}(\mathcal Q_m)=\lambda(\mathcal
Q_m). 
\label{toy1-tree}
\ee
At this point one could already run the LE parameters from $\mathcal Q_m$ to $\mathcal Q_{EW}$ and obtain the minimization condition. However the result would strongly depend on the choice of $\mathcal Q_m$. Indeed, since the LE and HE parameters run very differently, one would obtain different minimization conditions for different values of $\mathcal Q_m$ in Eq.~\eqref{toy1-tree}, even though the fundamental HE parameters would be kept fixed. This problem is alleviated by performing a one-loop matching. Hereafter we adopt the
$\overline{\rm MS}$ renormalization scheme to subtract the one-loop divergences.

In the present model there is no one-loop wave function renormalization and the tree level relation $\phi_{LE}(\mathcal Q_m)=\phi(\mathcal Q_m)$ is preserved at one loop.  The one-loop matching of the
other parameters can be obtained by matching the LE and the HE effective potentials where all parameters are at the scale $\mathcal Q_m$. We thus impose the relation
  \be
   V(\phi)= V_{LE}(\phi)  \,
   \eeq
with
 \bea
 V(\phi)&=&\Omega+\frac{1}{2}m^2\phi^2+\frac{1}{4!}\lambda\phi^4+\frac{1}{64\pi^2}\sum_{i=\phi,S} m_i^4\left(\log\frac{m_i^2}{\mathcal Q_m^2}-\frac{3}{2}\right) \,,\\
 V_{LE}(\phi)&=& \Omega_{LE}+\frac{1}{2} m_{LE}^2\phi^2+\frac{1}{4!}\lambda_{LE}\phi^4+\frac{1}{64\pi^2} m_{\phi_{LE}}^4\left(\log\frac{ m_{\phi_{LE}}^2}{\mathcal Q_m^2}-\frac{3}{2}\right)+\mathcal O(\phi^6/M^2)\,,\label{descomposicion}\nonumber
 \eea
 where $m_\phi^2=m^2+\frac{1}{2}\lambda\phi^2$, $m_S^2=M^2+h^2\phi^2$ and 
 $m_{\phi_{LE}}^2=m_{LE}^2+\frac{1}{2}\lambda_{LE}\phi^2$. This matching leads to
\begin{align}
\label{desacoplo}
 \Omega_{LE}(\mathcal Q_m)&= \left[\Omega(\mathcal Q_m)+\frac{M^4}{64\pi^2}\log\frac{M^2}{\mathcal Q_m^2}\right]-\frac{3 M^4}{128\pi^2}~,
\nonumber\\
 m_{LE}^2(\mathcal Q_m)&= \left[m^2(\mathcal Q_m) +\frac{h^2 M^2}{16\pi^2}\log\frac{M^2}{\mathcal Q_m^2}\right]-\frac{h^2 M^2}{16\pi^2}~ ,\\
 \lambda_{LE}(\mathcal Q_m)&=\left[\lambda(\mathcal Q_m) +\frac{12}{32\pi^2}h^4\log\frac{M^2}{\mathcal Q_m^2}\right]~,\nonumber
\end{align}
which holds up to two-loop corrections~\footnote{Notice that the
  difference between LE and HE parameters is at one-loop. Using the
  former or the latter in the one-loop contribution to the effective
  potential makes a difference only at the two-loop order.} and where
all parameters are evaluated at the scale $\mathcal Q_m$. In
particular due to the one-loop scale invariance of the HE and LE
effective potentials, the arbitrariness of $\mathcal Q_m$
in~\eqref{desacoplo} is guaranteed at one-loop level.  In fact, within
each parenthesis in \eqref{desacoplo} the logarithms compensate the
one-loop running of the parameters whose evolutions are given by the
$\beta$-functions
\begin{align}
\beta_{\Omega_{LE}}&=\frac{1}{32\pi^2} m_{LE}^4,\quad \beta_\Omega=\frac{1}{32\pi^2}(m^4+M^4)\ ,
\nonumber\\
\beta_{ m_{LE}^2}&=\frac{1}{8\pi^2}\lambda_{LE} m_{LE}^2,\quad \beta_{m^2}=\frac{1}{8\pi^2}(\lambda m^2+h^2 M^2)\ ,
\\
\beta_{\lambda_{LE}}&=\frac{6}{32\pi^2}\lambda_{LE}^2,\quad \beta_\lambda=\frac{6}{32\pi^2}(\lambda^2+4h^4)\nonumber \ .
\end{align}
This shows that in the one-loop approximation the running and matching procedures are commutative and the LE theory is independent of the decoupling scale.

The one-loop freedom in the choice of $\mathcal Q_m$ can be used to minimize the higher order corrections to (\ref{desacoplo}).  To this aim it is sensible to choose $\mathcal Q_m=M$. In this particular case the matching for $m_{LE}^2$ turns out to be
 \beq
 m_{LE}^2(M)=m^2(M)-\frac{h^2}{16\pi^2}\,M^2\ .
 \eeq
 The second term is the large threshold correction that, in a more realistic theory as the SM, would destabilize the electroweak vacuum and would introduce a hierarchy problem.

\subsection{Second toy model: scalars and fermions}
\label{2toy}
\noindent 
Further subtleties of the decoupling procedure can be described in a slightly more complicated scenario. We consider a HE theory consisting of a light real scalar $\phi$, a light Dirac fermion $\Psi$, and a number of heavy scalars $S_I$ with masses $M_I$ of similar magnitude:
\be
\mathcal
 L=-\Omega+\frac{1}{2}(\partial\phi)^2-\frac{m^2}{2}\phi^2-\frac{\lambda}{4!}\phi^4+\frac{1}{2}\sum_I\left[(\partial S_I)^2-(M_I^2 +h_I^2\phi^2)S_I^2\right]
 + i\bar
   \Psi\slashed \partial \Psi  - Y \phi\bar\Psi\Psi~. 
 \label{lagr1}
\ee
This theory can be described at LE by
\beq 
\mathcal
 L_{LE}=-\Omega_{LE}+\frac{1}{2}(\partial\phi_{LE})^2-\frac{m_{LE}^2}{2}\phi_{LE}^2-\frac{\lambda_{LE}}{4!}\phi_{LE}^4+ i\bar
   \Psi_{LE}\,\slashed \partial \Psi_{LE}  - Y_{LE} \phi_{LE}\bar\Psi_{LE}\,\Psi_{LE} 
\label{lagr2}
\eeq
and the dependence with respect to $\mathcal Q$ is implicit in Eqs.~\eqref{lagr1} and \eqref{lagr2}.

Since the $S_I$ fields do not acquire any VEV and their integration do not lead to any tree-level correction to the LE couplings, the tree-level matching is trivial. Instead the one-loop matching is less straightforward due, for instance, to the radiative corrections to the kinetic terms.  As we are interested in the one-loop electroweak-breaking conditions, we will focus on the effective
potential of the light scalar field. In particular we want to calculate the RG-improved one-loop matching at the scale $\mathcal Q_m$. To this aim we first run all parameters in \eqref{lagr1} and \eqref{lagr2} to $\mathcal Q_m$, and then we evaluate the LE and HE one-loop potentials. We thus absorb the kinetic-term radiative corrections~\footnote{In the rest of the paper we will consider only the one-loop RG-improved Coleman-Weinberg potential and wave function  corrections. We then simplify the notation by omitting the superscript ``(1)'' in the quantities $Z_i^{(1)}$ and $\Delta V^{(1)}$ introduced in section~\ref{decoupling}.} $(1-2Z_\phi(\mathcal Q_m))$ and $(1-2Z_{\phi_{LE}}(\mathcal Q_m))$ into field redefinitions of $\phi$ and $\phi_{LE}$ respectively, where $d[Z_i(\mathcal Q)]/d\log\mathcal Q=\gamma_i(\mathcal Q)$, the anomalous dimension at one-loop order.  The expansion of the HE RG-improved one-loop potential at the scale ${\mathcal Q_m}$ has to reproduce the LE one at the same scale. We hence impose the matching
 \beq
   V(\phi)= V_{LE}(\phi)  \,
   \eeq
with
  \bea
 V(\phi)&=&\Omega+\frac{1}{2}m^2(1+2Z_\phi)\phi^2+\frac{1}{4!}\lambda(1+4 Z_\phi)\phi^4
+\sum_{i=\phi,S_I,\Psi} n_i \frac{m_i^4}{64\pi^2}\left(\log\frac{m_i^2}{\mathcal Q_m^2}-\frac{3}{2}\right) \,,\nonumber\\
V_{LE}(\phi_{LE})  &=&\Omega_{LE}+\frac{1}{2}m_{LE}^2(1+2Z_{\phi_{LE}})\phi_{LE}^2+\frac{1}{4!}\lambda_{LE}(1+4 Z_{\phi_{LE}})\phi_{LE}^4\nonumber\\
&+&\sum_{i=\phi_{LE},\Psi_{LE}} n_i \frac{m_i^4}{64\pi^2}\left(\log\frac{m_i^2}{\mathcal Q_m^2}-\frac{3}{2}\right)+\mathcal O(\phi_{LE}^6/M^2)\,,\nonumber
 \eea  
 where $m_\phi^2=m^2+\frac{1}{2}\lambda\phi^2$, $m_{S_I}^2=M_I^2+h^2\phi^2$,
 $m_{\Psi}^2=Y^2\phi^2$,
 $m_{\phi_{LE}}^2=m_{LE}^2+\frac{1}{2}\lambda_{LE} \phi_{LE}^2$,
 $m_{\Psi_{LE}}^2=Y_{LE}^2\phi_{LE}^2$, and $n_i=1 (-4)$ for real
 bosons (Dirac fermions).  This implies
\begin{align}
\Omega_{LE}(\mathcal Q_m)&= \left[\Omega(\mathcal Q_m)+\sum_{I}\frac{M_I^4}{64\pi^2}\log\frac{M_I^2}{\mathcal Q_m^2} \right]-\sum_{I}\frac{3 M_I^4}{128\pi^2}~,
\nonumber\\
m_{LE}^2(\mathcal Q_m)&=\left[ m^2(\mathcal Q_m)[1+2Z_\phi(\mathcal Q_m)-2Z_{\phi_{LE}}(\mathcal Q_m)] +\sum_I\frac{h_I^2 M_I^2}{16\pi^2}\log\frac{M_I^2}{\mathcal Q_m^2} \right]-\sum_I\frac{h_I^2 M_I^2}{16\pi^2} ~ ,\nonumber\\
\lambda_{LE}(\mathcal Q_m)&=\lambda(\mathcal Q_m)[1+4Z_\phi(\mathcal Q_m)-4Z_{\phi_{LE}}(\mathcal Q_m)] +\sum_I\frac{12 h_I^4}{32\pi^2}\log\frac{M_I^2}{\mathcal Q_m^2}~.
\label{multidesacoplo}
\end{align}
where we are using the tree-level matching condition $Y_{LE}(\mathcal
Q_m)=Y(\mathcal Q_m)$~\footnote{The one-loop matching condition for
  the Yukawa coupling $Y$ can be obtained diagrammatically. For the
  purposes of the present paper we do not need to make it explicit.}.

In this toy model only light degrees of freedom generate $Z_\phi$ and $Z_{\phi_{LE}}$ at one loop via couplings that have trivial tree-level matching: it thus follows that $Z_\phi=Z_{\phi_{LE}}$~\footnote{Note that in scenarios (as the MSSM) where also some heavy fields contribute to $Z_\phi$, the $Z_\phi-Z_{\phi_{LE}}$ dependence on $\QQm$ is necessary to guarantee the scale independence of the one-loop matching conditions.}. Moreover, in view of minimizing higher loop corrections in the matching of $m_{LE}^2$, one can adopt the choice $\mathcal Q_m=\bar M$,
where $\bar M$ is defined as
 \beq
 \log \bar M^2=\frac{\sum_I h_I^2M_I^2\log (M_I^2)}{\sum_I h_I^2M_I^2} \,,
 \label{multiQ0}
\eeq
so that only the non-logarithmic one-loop contribution is left in the matching condition for $m_{LE}^2$: 
 \beq
\label{matchingMultiDesac}
 m_{LE}^2(\bar M)
 =m^2(\bar M) 
 -\sum_I\frac{h_I^2}{16\pi^2}\,M_I^2~.
\eeq
Of course the choice \eqref{multiQ0} is expected to reduce the higher order correction in the $m_{LE}^2$ matching only if the relation $\log\bar M^2\sim\log M_I^2$ occurs for all $I$'s. Otherwise, as already stressed in section~\ref{decoupling}, it is worth reiterating the decoupling procedure at each mass scale $M_I$.

\section{\sc The SM/MSSM matching}
\label{MSSM}
\noindent
In this section we use the MSSM one-loop RG-improved effective potential to determine the radiative corrections that can destabilize the electroweak breaking condition in such a model. The Higgs sector contains two doublets $H_1$ and $H_2$ where in our convention $H_2$ gives the mass to the top quark and $H_1$ to the bottom quark and tau lepton. The MSSM Higgs Lagrangian, including only the one-loop wave function renormalization for the Higgs doublets $Z_{h_i}$, can be written as
\be
\mathcal L_{MSSM}=\sum_i (1-2 Z_{h_i})|D_\mu H_i|^2-V(H_1,H_2)
\ee
where $V(H_1,H_2)$ is the tree-level MSSM Higgs potential. Then the one-loop RG-improved MSSM potential of the neutral Higgs fields $h_i=\textrm{Re}H_i^0$ ($i=1,2$) is
\bea
 V(h_1,h_2)&=&m_1^2(1+2Z_{h_1})h_1^2+m_2^2 (1+2Z_{h_2}) h_2^2-2
m_3^2 h_1 h_2 (1+Z_{h_1}+Z_{h_2})
\nonumber
\\
&+& \frac{g_Z^2}{2}[h_1^2(1+2Z_{h_2})-h_2^2(1+2Z_{h_2})]^2
+\Delta V_{MSSM}(h_1,h_2)\ , 
\label{V1_MSSM}
\eea
where $\Delta V_{MSSM}$ is the Coleman-Weinberg contribution generated by all fields of the MSSM and~\footnote{We are using conventionally for the electroweak gauge couplings the notation: $g^\prime\equiv g_Y$ and $g\equiv g_2$. $g_Y$ is related to the $U(1)$ gauge coupling $g_1$ in the $SU(5)$ normalization by $g_Y=\sqrt{\frac{3}{5}}g_1$.} 
\begin{equation}
g_Z^2\equiv (g_Y^2+g_2^2)/4,\quad m_1^2\equiv m_{H_1}^2+\mu^2,\quad m_2^2\equiv
m_{H_2}^2+\mu^2 .  
\label{algunasdef}
\end{equation}
The above equation is understood at an arbitrary renormalization scale $\mathcal Q$.
 
In view of the strong LHC bounds on the masses of supersymmetric
particles we match the MSSM with the SM at some high scale $\mathcal
Q_m$, say (multi-)TeV. To this aim we employ the effective potential
techniques adopted in the previous examples. For simplicity, we assume
all parameters to be real although the extension to cases with complex
parameters (and $CP$ violation) is straightforward.

Contrarily to the previous examples, the MSSM has two fields, in the
gauge eigenstate basis, that acquire VEVs. We then go to the mass
eigenstate basis to work out the matching (first at the tree-level,
then at one-loop).  The field rotation can be performed by neglecting
the $\mathcal O(v^2/\mathcal Q_m^2)$ electroweak-breaking contributions (i.e.~we proceed in the electroweak symmetry unbroken phase) since the $CP$-odd Higgs mass
$m_A$ is assumed to be much heavier than the electroweak scale. The
resulting potential can be matched to the SM potential whose one-loop
RG-improved expression is given by
\be 
V_{LE}(h_{LE})=-m_{LE}^2(1+2Z_{h_{LE}}) h_{LE}^2+\frac{\lambda_{LE}}{2}(1+4Z_{h_{LE}})
 h_{LE}^4 + \Delta V_{SM}(h_{LE})~ ,
\label{SMpot}
\ee
where $\Delta V_{SM}$ is the SM one-loop RG-improved Coleman-Weinberg potential in the presence of the background field $h_{LE}=\textrm{Re}\,H_{LE}^0$ (with $H_{LE}$ being the SM Higgs doublet). In Eq.~\eqref{SMpot} and hereafter the effective higher-order operators, which are small due the large hierarchy between heavy and light fields, are neglected.

\subsection{Tree-level matching}
\label{tree-level-rotation}
\noindent
In order to derive the RG-improved tree-level matching we will focus on the quadratic part of the tree-level MSSM potential, $V^{(0)}(h_1,h_2)$, which can be extracted from (\ref{V1_MSSM}).
At the matching scale $\mathcal Q_m$ we thus obtain
\bea
 V^{(0)}(h_1,h_2)&=&m_1^2 h_1^2+m_2^2 h_2^2-2
m_3^2 h_1 h_2 + \frac{g_Z^2}{2}(h_1^2-h_2^2)^2  \ . 
\label{V0_MSSM}
\eea
The potential $V^{(0)}(h_1,h_2)$ at $\mathcal Q_m$ can be rewritten in the mass eigenstate basis as
\be 
V^{(0)}(h,H)=-m^2h^2+ m_H^2 H^2+ \cdots \ 
\label{V0_SM} 
\ee
with $m^2 \ll m_H^2=m_1^2+m_2^2+m^2$. This field transformation is achieved by the rotation
\be
\left(\begin{matrix}
h_1 \\ h_2
\end{matrix}\right) 
=
R_\beta\left(\begin{matrix}
h \\ H
\end{matrix}\right) \ ,\quad
R_\beta= 
\left(\begin{matrix}
\cos\beta & -\sin \beta\\
\sin\beta & \cos\beta
\end{matrix}\right)
\label{Rbeta1}
\ee
such that
\bea
(h_1,h_2)\left(\begin{matrix}
m_1^2 & -m_3^2\\
-m_3^2 & m_2^2
\end{matrix}\right)
\left(\begin{matrix}
h_1 \\ h_2
\end{matrix}\right) = 
(h,H)\left(\begin{matrix}
-m^2 &0\\
0 & m^2_H
\end{matrix}\right)
\left(\begin{matrix}
h \\ H
\end{matrix}\right)\ .
\label{Rbeta2}
\eea
Note that Eqs.~\eqref{Rbeta1} and \eqref{Rbeta2} are equivalent to require
\begin{align}
m_3^4&=(m_1^2+m^2)(m_2^2+m^2)~,\label{eq:m3tree}\\
\tan 2\beta&=\frac{2 m_3^2}{m_2^2-m_1^2}
\label{eq:gammatree}
\end{align}
or, alternatively
\begin{align}
m_2^2&=m_3^2 /\tan\beta-m^2\nonumber ~,\\
m_1^2&=m_3^2 \tan\beta-m^2 ~, \label{matching1}
\end{align}
with $\tan\beta\geq 1$, as we are assuming $m_3^2>0$ and $m_1^2\geq m_2^2$.  In particular since we are in the decoupling limit $m^2 \ll m_H^2$, our definition of $\tan\beta$ coincides with the MSSM one, $\tan\beta=v_2/v_1$ where $v_i\equiv\langle h_i\rangle$.  Combining the above relations we also have the explicit expression for the light mass eigenstate 
\bea m^2&=&- m_1^2
\cos^2\beta- m_2^2\sin^2\beta+ m_3^2\sin 2\beta ~.
\label{m2tree}
\eea

In the mass eigenstate basis $(h,H)$ it is easy to obtain the tree-level matching to the SM. If one extracts the $V^{(0)}_{LE}(h_{LE})$ part from the LE one-loop potential of $h_{LE}$, Eq.~(\ref{SMpot}),  and matches it to $V^{(0)}(h,H)$ at $\mathcal Q_m$, one obtains
\be
h_{LE}(\mathcal Q_m)=h(\mathcal Q_m) ,\ m_{LE}(\mathcal
Q_m)=m(\mathcal Q_m) ,\ \lambda_{LE}(\mathcal Q_m)=g_Z^2(\mathcal Q_m) \cos^2 2\beta~\ {\rm[tree~level]}\ .
\label{treeLevel}
\ee
Similarly the matching of the LE and HE Yukawa interactions at $\mathcal Q_m$ yields
 \be 
 y_{t}(\mathcal Q_m)
= Y_{t}(\mathcal Q_m) \sin\beta\ ,\quad 
y_{b,\tau}(\mathcal Q_m)
=Y_{b,\tau}(\mathcal Q_m) \cos\beta~,\qquad {\rm[tree~level]}\ \label{treeLevel2}
 \ee
where $y_{t,\tau,b}$ and $Y_{t,\tau,b}$ are respectively the top quark, tau lepton and bottom quark Yukawa couplings in the SM and in the MSSM. 

Of course there might already be a problem at tree level: the right
hand side of Eq.~(\ref{m2tree}) is a linear combination of potentially
large masses squared while the left hand side is a mass squared which
is required to be at the electroweak scale. This fine-tuning is
essentially equivalent to that in Eq.~(\ref{eq:m3tree}). This is the
main naturalness problem in the MSSM. This problem cannot be tackled
unless we know the (fundamental) theory responsible for triggering
supersymmetry breaking at the high scale $\mathcal M$ in the hidden
sector and dictating the size of the supersymmetry breaking parameters
in the observable sector. The FP
solution~\cite{Feng:1999zg,Delgado:2014vha} just uncovers the
functional relationships between fundamental parameters at the high
scale $\mathcal M$ for which the naturalness problem is
circumvented. However even if we accept that the fundamental theory
might provide a solution to the tree-level stability we still have to
worry for loop corrections, e.g.~in the effective theory as those
computed in Ref.~\cite{Veltman:1980mj}.  The matching including one
loop corrections will be done in the next section.

\subsection{One-loop level matching}
\noindent
We now proceed with the one-loop matching. Again we want to work in the mass eigenstate basis. We then impose the tree-level matching conditions (\ref{matching1}) in the one-loop term $\Delta V_{MSSM}$ of \eqref{V1_MSSM}, and expand $\Delta V_{MSSM}$. Such expansion produces some new quadratic contributions that we absorb as
\be
V(h_1,h_2)=\widetilde m_1^2  h_1^2+\widetilde m_2^2h_2^2-2 \widetilde m_3^2 h_1 h_2 + \cdots
\label{MSSMpot2}
\ee
where
\bea
&\widetilde m_i^2=m_i^2+2Z_{h_i}m_i^2+\Delta m_i^2~,\qquad
&\Delta m_{i}^2=\left.\frac{\partial \Delta V_{{MSSM}}}{\partial h_i^2}\right|_{h_i=0}~ ,
\nonumber\\
&\widetilde m_3^2=m_3^2+(Z_{h_1}+Z_{h_2})m_3^2+\Delta m_3^2~,\qquad
& \Delta m_{3}^2=-\frac{1}{2}\left.\frac{\partial \Delta V_{{MSSM}}}{\partial h_1 h_2}\right|_{h_i=0} ~,
 \label{aqui}
\eea
with $i=1,2$. As previously done for the tree-level matching, we diagonalize the quadratic potential (\ref{MSSMpot2}) by a rotation $R_\beta$ (whose angle differs from that of
section~\ref{tree-level-rotation} although for notational simplicity we are keeping the same notation for both) leading to a light mass eigenstate $h$ with squared mass $-\widetilde m^2$ and a heavy eigenstate $H$ with squared mass $\widetilde m_H^2= \widetilde m_1^2+\widetilde m_2^2+\widetilde m^2$, where 
\bea
\label{Dm2}
\widetilde m^2&=& m^2+2 m^2 Z_h+\Delta m^2 \, , \nonumber \\
m^2&=&- m_1^2 \cos^2\beta- m_2^2\sin^2\beta+ m_3^2\sin 2\beta \, , \\
\Delta m^2&=&-\Delta m_1^2 \cos^2\beta-\Delta m_2^2 \sin^2\beta+\Delta
m_3^2 \sin 2\beta \nonumber~ ,
\eea
and $Z_h\equiv\cos^2\beta Z_{h_1}+\sin^2\beta Z_{h_2}$ is the wave function renormalization in the MSSM for the mass eigenstate $h$~\footnote{Notice that the identity $Z_h m^2\equiv Z_{h_1}m_1^2\cos^2\beta +Z_{h_2}m_2^2\sin^2\beta -(Z_{h_1}+Z_{h_2})m_3^2\sin 2\beta $ is obtained after using the tree-level matching conditions, Eq.~(\ref{matching1}), on the masses and mixing angle, as required by the fact that the wave function renormalization is already a one-loop effect.}.
This diagonalization requires
\begin{align}
  \widetilde m_2^2&=\widetilde m_3^2/ \tan\beta- \widetilde m^2  ~,\nonumber\\
  \widetilde m_1^2&=\widetilde m_3^2\tan\beta-\widetilde m^2
  \ ,
 \label{matchingrad}
\end{align}
which can be used to express, as it is customary in the MSSM,
$\tan\beta$ and the lightest eigenvalue $-\widetilde m^2$ as
functions of the fundamental parameters:
\bea
\widetilde m^2         
&=&-\widetilde m_1^2 \cos^2\beta-\widetilde m_2^2\sin^2\beta+\widetilde m_3^2\sin 2\beta\ , \nonumber \\
& & \tan 2\beta=\frac{2\, \widetilde m_3^2}{\widetilde m_2^2-\widetilde m_1^2}\ . 
\label{output}
\eea

In order to perform the complete one-loop matching, we would need to consider $\Delta V_{SM}$. As the light (i.e.~SM) fields provide the same contributions to $\Delta V_{SM}$ and $\Delta V_{MSSM}$ modulo the tree-level matching of Yukawa couplings in Eq.~(\ref{treeLevel2}) (cf.~also section~\ref{radiative}), we can proceed by taking into account only the heavy non-SM fields in $\Delta m_1^2$, $\Delta m_2^2$ and $\Delta m_3^2$. Then the one-loop RG-improved matching of the quadratic term in the HE and LE theories turns out to be
\bea
\label{deltam}
m^2_{LE}(\mathcal Q_m)&=&m^2(\mathcal Q_m) (1+2\Delta Z_h(\mathcal Q_m))+\Delta m^2(\mathcal Q_m) ~ ,  
\eea
where
\bea    
\Delta Z_h(\mathcal Q_m)&=&Z_h(\mathcal Q_m)-Z_{h_{LE}}(\mathcal Q_m)~ .
\label{deltaZ}  
\eea
We hence stress that the requirement $m^2(\mathcal Q_m)\sim \mathcal Q^2_{EW}$ is not sufficient to guarantee sensible electroweak breaking conditions as these could be destabilized by $\Delta m^2$ and $\Delta Z_h$.

\section{\sc $\Delta m^2$ in the MSSM with high susy-breaking scale}
\label{radiative}
\noindent
We will now explain the main lines to determine $\Delta m^2$.  We
remind that all the MSSM particles, except the SM ones, are assumed to be
heavy. The one-loop potential
$\Delta V_{MSSM}(h_1,h_2)$ can be split into two separated terms: one term contains contributions from
the Higgs sector $A,H,H^\pm$ and the SM fields, and the second one does it
from the field superpartners $\tilde f$ (with $\tilde f$ representing
the whole list of squarks, sleptons, charginos and neutralinos).  Due
to the triviality of the tree-level matching conditions
\eqref{treeLevel} and \eqref{treeLevel2}, it is easy to see as we already noticed that the
light fields provide the same contribution to $\Delta V_{MSSM}(h)$
and to $\Delta V_{SM}(h_{LE})$.  Because of this property the
correction $\Delta m^2$ can be calculated as in Eq.~\eqref{Dm2} with
\bea
\label{threshEFF}
\Delta m_{I}^2 &=& \sum_{r=\tilde f,A,H,H^\pm}  \Delta m_{I,r}^2\, , \\
 \Delta m_{i,r}^2  = \left. \frac{\partial \Delta V_{{MSSM,r}}}{\partial h_i^2 } \right|_{h_i=0} && 
  \Delta m_{3,r}^2  = \left. -\frac{1}{2} \frac{\partial \Delta V_{{MSSM,r}}}{\partial h_1 h_2 } \right|_{h_i=0}~,
 \eea 
where $ I=(i,3)$, $i=1,2$ and  $\Delta V_{MSSM,r}$ is the MSSM one-loop potential generated by the field $r$. 

The explicit form of $\Delta m_{I,r}^2$ depends on the renormalization scheme. In the $\overline{\rm MS}$ scheme (or equivalently at this level $\overline{\rm DR}$), for which $\mathcal C= (3/2) {\bf 1}$ in Eq.~\eqref{CW} for scalars and fermions, it results
\bea
\Delta m_{i,r}^2 =\left.\frac{n_r }{32\pi^2} \frac{\partial m_r^2}{\partial h_i^2} G(m_{r}^2)\right|_{h_i=0}
\ ,\,\,\,\,\, 
\Delta m_{3,r}^2 =\left. -\frac{n_r }{64\pi^2} \frac{\partial m_{r}^2}{\partial h_1h_2} G(m_{r}^2)\right|_{h_i=0}~ ,  \label{Dmsqs}
\eea
where $n_r$ stands for the number of degrees of freedom of the particle $r$ and is positive (negative) for bosons (fermions), while the function $G(x^2)$ is defined as
\bea
G(x^2)\equiv x^2\left( \log\frac{x^2}{\mathcal Q_m^2}-1  \right)\, .
\label{Gmsbar}
\eea
For simplicity we determine $\Delta m^2$ by neglecting the corrections
coming from first and second generations of squarks and sleptons
(the expressions are however fully general and the first two generation
sfermions could be easily included). Details of the calculation are
furnished in appendix~\ref{appendixA}.  Here we only report on the final
result: 
\begin{align}
-\Delta m^2&=\frac{1}{32\pi^2}\Big\{
6y_t^2[G(m_Q^2)+G(m_U^2)]+6y_b^2[G(m_Q^2)+G(m_D^2)]+2y_\tau^2[G(m_L^2)+G(m_E^2)]\nonumber\\
&+6 y_t^2 X_t^2\frac{G(m_Q^2)-G(m_U^2)}{m_Q^2-m_U^2}+6 y_b^2 X_b^2\frac{G(m_Q^2)-G(m_D^2)}{m_Q^2-m_D^2}+2 y_\tau^2 X_\tau^2\frac{G(m_L^2)-G(m_E^2)}{m_L^2-m_E^2}\nonumber\\
&-g_Y^2 \left(G(m_Q^2)-2 G(m_U^2)+G(m_D^2)-G(m_L^2)+G(m_E^2)  \right)\cos2\beta\nonumber\\
&-6 g_2^2\frac{M_2^2G(M_2^2)-\mu^2G(\mu^2)}{M_2^2-\mu^2}-2 g_Y^2\frac{M_1^2G(M_1^2)-\mu^2G(\mu^2)}{M_1^2-\mu^2}\nonumber\\
&-\left(12 g_2^2 M_2\mu\frac{G(M_2^2)-G(\mu^2)}{M_2^2-\mu^2}+4 g_Y^2 M_1\mu\frac{G(M_1^2)-G(\mu^2)}{M_1^2-\mu^2}   \right)\sin\beta\cos\beta\nonumber\\
&+G(m_H^2)\left(-6 g_Z^2 \cos^22\beta+2 g_Z^2+g_2^2  \right)\Big\}\ ,
\label{Deltam2}
\end{align}
where the soft breaking terms $X_{t,b,\tau}$ are defined as
\be
X_t= A_t -\frac{\mu}{\tan\beta} \, ,\quad \, X_{b,\tau}= A_{b,\tau} - \mu \tan\beta    \,\,.
\ee
In Eq.~\eqref{Deltam2} the first two lines correspond to the
  contribution from sfermions, the third line the Fayet-Iliopoulos
  contribution from scalars, the forth and fifth lines the contribution
  from charginos and neutralinos, and the last line the contribution
  from the heavy scalar, the pseudoscalar and charged Higgses. Notice
  that all supersymmetric parameters are defined at the scale
  $\mathcal Q_m$.

We therefore conclude that in the MSSM with heavy non-SM particles,
the Higgs sector at LE appears like the one of the SM where the Higgs
quadratic parameter at the scale $\mathcal Q_m$ is given by the
relation of Eq.~\eqref{deltam} with $m^2$ and $\Delta m^2$ as in
Eqs.~(\ref{Dm2}) and (\ref{Deltam2}) and $\tan\beta$ given by
Eq.~(\ref{output}). Moreover the explicit expression of $\Delta
Z_h$~\footnote{In general $\Delta Z_h$ consists of two terms: one
  depending on the renormalization scale and proportional to the
  anomalous dimension difference $\gamma_h-\gamma_{h_{LE}}$; and a
  second one leading to a one-loop scale-independent difference
  between the LE and HE parameters (see e.g.~\cite{Carena:2008rt}).}
is not required in first approximation as we will see in
section~\ref{quadratic}. Finally, it is worth noting that, by
construction, in the considered heavy MSSM scenario the electroweak
breaking condition at $\mathcal Q=\mathcal Q_{EW}$ can be evaluated in
the LE theory (avoiding large logarithms) with no one-loop dependence
on the choice of $\mathcal Q_m$ (cf.~appendix~\ref{appendixB}).

\section{\sc The stability of the SM effective theory}
\label{quadratic}
\noindent

As reminded in section~\ref{MSSM}, even in the case that the
fundamental theory naturally leads to $m^2$ of the order of the
electroweak scale, we still have to worry about the destabilization
and unnaturalness due to radiative corrections. In this section we
sketch some relationships that would help to not destabilize the
vacuum.
These can be deduced after evaluating the size of the radiative
corrections at $\QQm$ and forcing them to be in the ballpark of 
$m^2(\mathcal Q_m)$.

Some preliminary observations are in order here:
\begin{itemize}
\item In general $m^2(\mathcal Q)$ has a strong scale dependence and
  its value can span several order of magnitudes. Its (one-loop)
  radiative correction is given by $\Delta m^2(\mathcal
  Q)+2m^2(\mathcal Q)\Delta Z_h(\mathcal Q)$, which is also strongly
  scale dependent. Hence claiming $m^2=\mathcal O(100\ \rm{GeV})^2$ 
  makes sense only at a specific running scale and, concerning
  the hierarchy problem, the result is satisfactory only if at the
  same scale the radiative corrections contain no large logarithms (to
  keep perturbation theory trustable) and are of the order of the
  electroweak scale or below (to not destabilize the tree level
  result). In this section we assume that the supersymmetry breaking
  theory yields $m^2=\mathcal O(100\,\rm{GeV})^2$ at the specific
  scale $\QQm$.
\item The strong (one-loop) scale dependences of $m^2(\mathcal Q)$ and
  its radiative correction have opposite signs and almost cancel out.
  Indeed the $\mathcal Q$-dependence of these two terms is
  equivalent to the one of $m^2_{LE}(\mathcal Q)$
  [cf.~Eq.~\eqref{deltam}], which amounts to the $\beta$-function of
  the SM (cf.~appendix~\ref{appendixB}) and is thus negligible for our
  purposes~\footnote{We checked numerically that in the SM the
    quadratic term changes by about $\sim$10\% for a running from the
    electroweak to the Planck scale.}.
\item 
Heavy particles have masses well above $m^2(\QQm)$. This implies that the wave function correction $2m^2(\mathcal Q_m)\Delta Z_h(\mathcal Q_m)$ can be neglected in comparison to $\Delta m^2(\mathcal Q_m)$.
\end{itemize}

Therefore within the above approximations the stability of the
electroweak breaking conditions under radiative corrections can be
evaluated by means of the ``stability parameter''
\be
\mathcal S=\left|\frac{\Delta m^2(\mathcal Q_m)}{m^2_{LE}( \mathcal Q_m)}\right|
\label{stability}~,
\ee
where the running of $m^2_{LE}$ between $\mathcal Q_{EW}$ and $\QQm$
can be neglected. We now proceed determining $\mathcal S$ in some
specific scenarios.

\subsection{\sc Degenerate case}
\label{degener}
\noindent
We first consider the simplest MSSM scenario where all mass parameters
are degenerate at some common value $M$:
\beq
m_Q=m_U=m_D=m_L=m_E=M_1=M_2=\mu=m_H\equiv M~.
\eeq
In this case all radiative corrections depend on the single logarithm
$\log M^2/\mathcal Q_m^2$ and the simplest choice for the decoupling
scale is obviously $\mathcal Q_m=M$.  Eq.~\eqref{Deltam2} thus
yields
\begin{equation}
\Delta m^2(M)=\frac{M^2}{32\pi^2} \left\{ 12 y_t^2+12 y_b^2+4y_\tau^2-6\lambda -\frac{3}{2}g_Y^2-\frac{9}{2}g_2^2\right\}\ .
\label{VC}
\end{equation}
%
%

Eq.~(\ref{VC}) reproduces the SM Higgs mass quadratic divergence
obtained in the case that the SM has a cutoff $\Lambda\equiv M$
\cite{Veltman:1980mj, Einhorn:1992um}: the first three terms
correspond to the quadratic divergences coming from exchange of top,
bottom and tau fermions [with masses
$m_{t,b,\tau}(h_{LE})=y_{t,b,\tau} h_{LE}$], the fourth term
corresponds to divergences coming from exchange of the SM Higgs [with
mass $m_h^2(h_{LE})=-m^2+3\lambda h_{LE}^2$] and neutral and charged
Goldstone bosons [with masses $m_\chi^2(h_{LE})=-m^2+\lambda
h_{LE}^2$], while the last two terms are equivalent to those due to
exchange of $W$ and $Z$ gauge bosons [with masses $m_W^2(h_{LE})=g_2^2
h_{LE}^2/2$ and $m_Z^2(h_{LE})=(g_Y^2+g_2^2)h_{LE}^2/2$].  In fact
Eq.~(\ref{VC}) can be written as a function of SM running masses as
\beq
\Delta m^2(M)=\frac{M^2}{32\pi^2v^2}\left\{ \sum_{f} n_{f} m_{f}^2-3m_h^2-3m_Z^2-6m_W^2 
\right\} \  ,
\label{VC1}
\eeq
where $n_f$ is the number of degrees of freedom of the fermion $f$.

\begin{figure}[htb]
\vskip .5cm 
 \begin{center}
 \includegraphics[width=8cm]{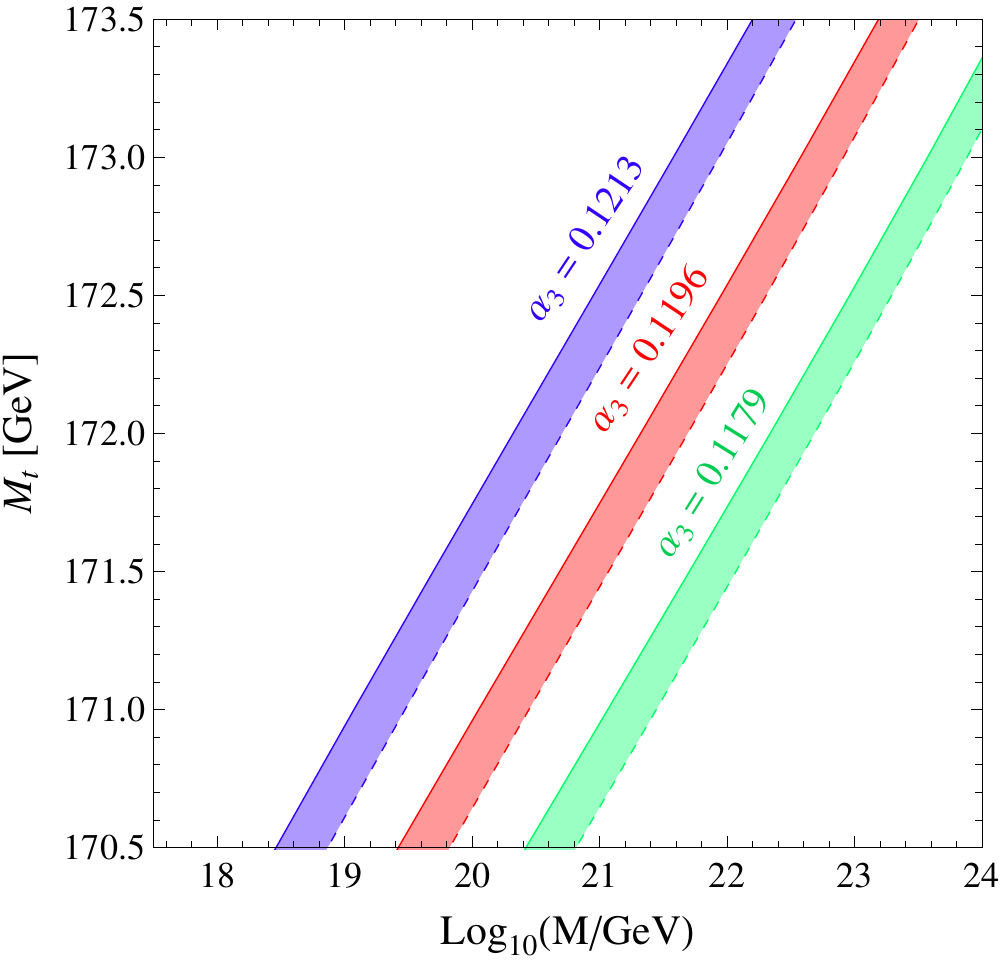}     
  \includegraphics[width=7.8cm]{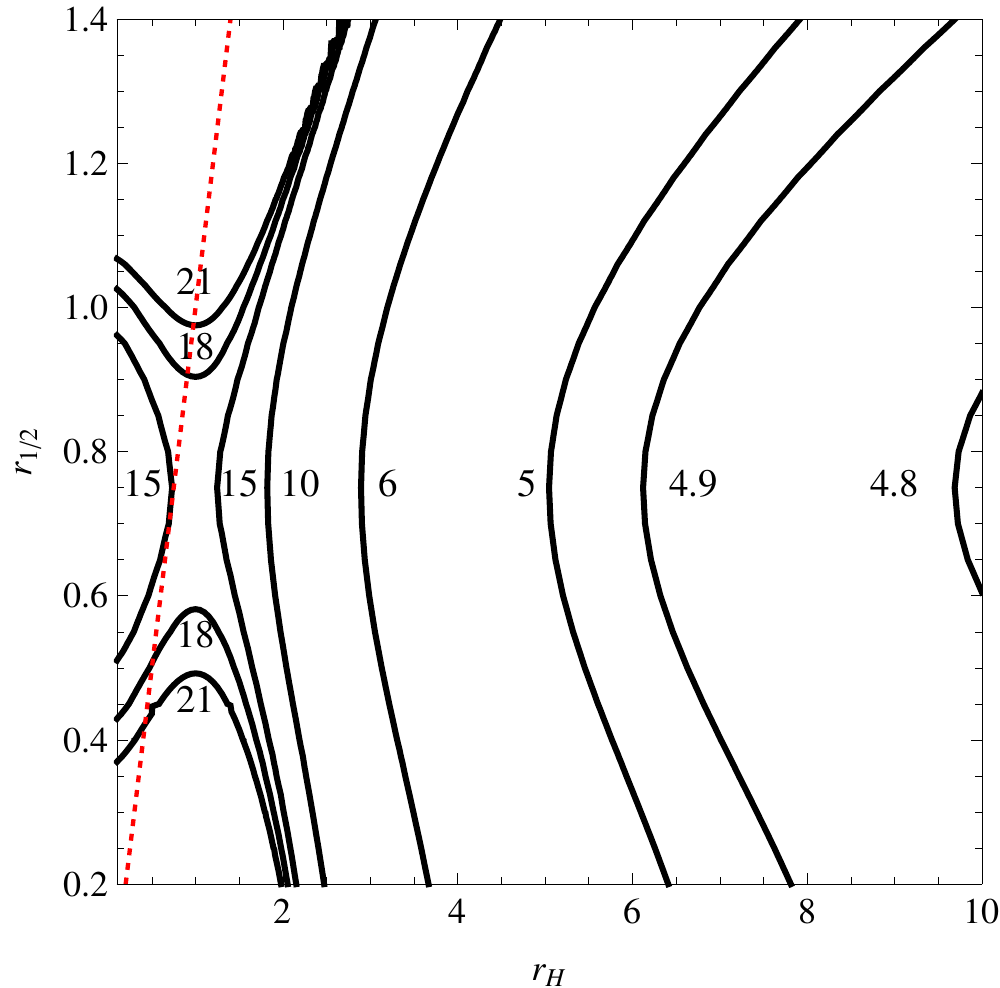} 
 \end{center}
 \caption{\it\baselineskip=15 pt Left panel: Contour lines of
   $\mathcal S=0$ for different values of $\alpha_3(m_Z)$ in the
   degenerate case of section~\ref{degener}. The solid (dashed) lines
   are obtained taking $m_h=126\,(125)$\,GeV. Right panel: Contour
   lines of $\log_{10}M_0$ as a function of $r_H$ and $r_{1/2}$
   leading to $\mathcal S=0 $ in the non-degenerate case of section
   \ref{nondegener}. We choose $m_h=125.1 \,\rm{GeV}$, $m_t=172.5\,
   \rm{GeV}$, $\alpha_3(m_Z)=0.1196$ and $\tan \beta=1$ for
   definiteness. Along the dotted straight line ($r_{1/2}=r_H$)  the
     quadratic corrections can be understood as a generalization of
     the Veltman condition.}
\label{fig-Vdeg}
\vskip .1 cm
\end{figure}

As it is well known, for experimental values of the SM masses the
requirement $\Delta m^2(M)=0$ in Eq.~\eqref{VC1}, usually dubbed
Veltman condition~\cite{Veltman:1980mj}, is not fulfilled at weak
scales but at Planckian scales~\cite{Masina:2013wja}. The value of
this high scale is quantified in the left panel of Fig.~\ref{fig-Vdeg}
where the contour lines of $\Delta m^2(M)=0$ (or equivalently
$\mathcal S=0$) are plotted in the plane $(\log_{10}M/{\rm GeV}, M_t)$
(where $M_t$ is the top quark pole mass), for different values of
$m_h$ and $\alpha_3(m_Z)$. The plot has been obtained by using the RG
equations of the SM parameters appearing in Eq.~(\ref{VC}) at the NNLO
(as done e.g.~in~\cite{Masina:2012tz}).

\subsection{\sc A simple non-degenerate case}
\label{nondegener}
\noindent
A simple non-degenerate case is the scenario where at the scale $\QQm$ the sfermion, electroweakino and Higgs sectors have each one a respective common mass $M_0$, $M_{1/2}$ and $M_H$ as:
\beq
m_Q=m_U=m_D=m_L=m_E\equiv M_0,\quad M_1=M_2=\mu\equiv M_{1/2},\quad m_H\equiv M_H\,\,.
\eeq
In this case we can express $\Delta m^2(\mathcal Q_m)$ as
\be
\Delta m^2(\mathcal Q_m) = \Delta_f m^2(\mathcal Q_m) +   \Delta_\ell m^2(\mathcal Q_m) 
\label{descomposicion} 
\ee
with~\footnote{The decomposition (\ref{descomposicion}) separates the
  one-loop scale independent contribution from the dependent one,
  $\Delta_f m^2(\mathcal Q_m)$ and $\Delta_\ell m^2(\mathcal Q_m)$
  respectively. This separation is defined up to an arbitrary one-loop
  scale independent quantity, but our results do not depend on the
  convention we are choosing.  }
\bea
\Delta_f m^2(\mathcal Q_m)&=&
\frac{1}{32\pi^2}\left\{
(12 y_t^2+12 y_b^2+4 y_\tau^2) M_0^2-4 G M_{1/2}^2-(6\lambda-G)M_H^2 
\right\} ~,
\label{VCgen}\\
\Delta_\ell m^2(\mathcal Q_m)&=& \frac{1}{32\pi^2}
\bigg\{-
\left[(12 y_t^2+12 y_b^2+4 y_\tau^2) M_0^2 + (6 y_t^2 X_t^2+6 y_b^2 X_b^2+ 2 y_\tau^2 X^2_\tau)\right]\log\frac{M_0^2}{Q_m^2}
\nonumber\\
 &+& 4 G M_{1/2}^2 f_\beta \log\frac{M_{1/2}^2}{Q_m^2} +  (6\lambda-G)M_H^2 \log\frac{M_H^2}{Q_m^2}\bigg\} ~,
\label{VCgenLoop}
\eea
where $G=(g_Y^2+3 g_2^2)/2$, $f_\beta=2+\sin 2\beta$ and all parameters are understood at the
scale $\QQm$. For definiteness we take $\QQm=M_{0}$ hereafter. It
follows
\bea
\Delta_f m^2(\mathcal Q_m)&=& \frac{M_{0}^2}{32\pi^2v^2}\Big\{ \sum_f n_f m_f^2\, -
4(2m_W^2+m_Z^2)r_{1/2}^2-(3m_h^2-2m_W^2-m_Z^2)r_H^2
\Big\}~,~~~~~~~~~
\label{eq-ndDeltam2}\\
\Delta_\ell m^2(\mathcal Q_m)&=& \frac{M_{0}^2}{32\pi^2v^2}
\Big\{
4 f_\beta (2m_W^2+m_Z^2)r_{1/2}^2\log r_{1/2}^2 +(3m_h^2-2m_W^2-m_Z^2)r_H^2 \log r_H^2
\Big\}~,~ \nonumber\\
\label{eq-ndDeltam2Loop}
\eea
where $r_H=M_H/M_{0}$, $r_{1/2}=M_{1/2}/M_{0}$ and all masses are running.

For $r_H=r_{1/2}$ the right hand side of Eq.~\eqref{eq-ndDeltam2}
coincides with the quadratic divergences that one derives from the SM
effective potential by using as regulators the cutoff $\Lambda_F=M_0$
for fermions and the cutoff $\Lambda_B=M_{1/2}=M_H$ for
bosons~\footnote{A similar interpretation is not clear for
  $M_{1/2}\neq M_H$ due to the fact that both the heavy (neutral and
  charged) Higgs sector and charginos and neutralinos contribute to
  the generalized Veltman condition with terms proportional to squared
  gauge couplings.}. In this case the requirement $\Delta_f m^2=0$ can
be interpreted as a generalization of the Veltman condition expressing
the Higgs quadratic divergence in terms of SM parameters and
cutoffs~\footnote{In particular, for some choices of $\Lambda_B$ and
$\Lambda_F$ at the TeV scale, the condition $\Delta_f m^2=0$ can be
achieved at low energies.}.  However also the logarithmic
corrections are sizeable and contribute to destabilizing the potential. With the
constraint $r_H=r_{1/2}$, $\Delta_\ell m^2$ is negligible only when
$r_H=r_{1/2}\approx 1$. Correspondingly the condition $\Delta m^2=0$
with cutoffs at the TeV scale is not possible anymore.

However, for $r_H\neq r_{1/2}$ the stability under perturbative
corrections can still be fulfilled at scales of $\mathcal O(100 $ TeV)
if one requires the cancellation of the total correction
$\Delta m^2$. This can be seen in the right panel of
Fig.~\ref{fig-Vdeg} where the constraints on $r_H$ and $r_{1/2}$
leading to $\mathcal S =0$ (solid curves) are plotted for several
values of $M_0$. The figure is obtained using
Eqs.~\eqref{eq-ndDeltam2} and \eqref{eq-ndDeltam2Loop} with
$\tan\beta=5$, and running
the SM parameter at NNLO with boundary conditions $m_h=125.1$\,GeV,
$M_t=172\,$GeV and $\alpha_3=0.1196$ at the electroweak scale as
explained in Ref.~\cite{Masina:2013wja}. The former case, $r_H=r_{1/2}$, is exhibited as the dotted straight line in the right panel of Fig.~\ref{fig-Vdeg} where we can see that stability can be achieved for sub-Planckian scales.
  
\subsection{\sc General soft breaking terms}
\noindent
In principle one expects that all soft breaking parameters will be different at the scale $\QQm$. In this general case the finite threshold contribution to the SM Higgs mass parameter is given by  Eq.~\eqref{Deltam2}.
%
%
The required condition for keeping $\Delta m^2$ at the order of the
electroweak scale is then a hypersurface in the multidimensional space
of supersymmetric parameters $(m_Q, m_U, m_D, m_L, m_E, A_t, A_b,
A_\tau, m_1, m_2, \mu, M_1, M_2)$. In the next section some of
  these hypersurfaces are analyzed numerically for the case of
  negligible Fayet-Iliopoulos (FI) contribution~\footnote{We remind
  that the FI contribution is a RG invariant. Our assumption is thus
  valid only if the FI term is zero at the scale of supersymmetry
  breaking transmission. Otherwise it should be taken into account
  although its (tiny) contribution should not change the qualitative
  conclusions.}.
\\[.1cm]

\noindent
Before closing this section a couple of comments are in order. In this
section we have assumed $m^2(\QQm)$ at the electroweak scale without
specifing the origin of such a value.
%
%
The main ideas to naturally produce $m^2(\mathcal Q_m)$ at the
electroweak scale are twofold: 
\begin{itemize}
\item
If, \textit{i)} the whole MSSM Higgs sector is at the electroweak scale (in which case the LE effective theory is not the SM but a two
  Higgs doublet model), and \textit{ii)} the masses of the
  supersymmetric partners are in the low TeV region (i.e.~$\QQm \not\gg
  \mathcal Q_{EW}$), then $m^2(\QQm )$ is at the electroweak
  scale. Moreover the requirement $\left|\Delta m^2(\QQm)\right|
  \lesssim \mathcal O(100\, {\rm GeV})^2$ is also automatically
  satisfied. This parameter configuration might be excluded soon by
  the LHC lower bounds on heavy Higgs and superpartner masses and we
  will not further discuss it.
\item For $\tan\beta\gg 1$ the squared mass $m^2(\QQm)$ is similar to
  $m_2^2(\QQm)$ [cf.~Eq.~\eqref{Dm2}]. The value of the latter is
  naturally small in the FP parameter region of the MSSM. This
  possibility has been broadly studied in the
  literature~\cite{Feng:1999zg} although the modifications due to $\Delta
  m^2$ have been overlooked.
\end{itemize}

In the next section we will analyze the effects that the one-loop
corrections have on the FP solution.

\section{\sc Numerical results: focus point solutions}
\label{numerical}\noindent
In this section we concentrate on the FP solution including the
one-loop radiative correction $\Delta m^2$. We stress that till now we
have expressed $m^2(\QQm)$ and $\Delta m^2(\QQm)$ as functions of
supersymmetric parameters evaluated at the scale $\mathcal
Q_m$. However, in view of a more fundamental supersymmetry-breaking
description, $m^2$ and $\Delta m^2$ should be re-expressed in terms of
the supersymmetric parameters evaluated at the messenger scale $\mathcal M$ at which
supersymmetry breaking is transmitted to the observable sector. As,
depending on the supersymmetry breaking model, some of the parameters
can unify at the messenger scale, the number of independent parameters at
the scale $\mathcal M$ can be smaller (than what would show up at the
scale $\QQm$). So, the required relation to keep $\Delta m^2$ and
$m^2(\QQm)$ of the order of the electroweak scale is simpler, and
might appear more natural.  This scenario is considered in this
section assuming, at the scale $\mathcal M$, as a simple example the case~\footnote{Other cases can be obviously considered along similar lines. Here we just present the case of Eq.~(\ref{highscale}) dubbed as NUHM1~\cite{AbdusSalam:2011fc}.}
\be
m_Q(\mathcal M)=m_U(\mathcal M)=M_0,\quad m_1(\mathcal M)=m_2(\mathcal M)=M_H,\quad M_a(\mathcal M)=M_{1/2}~.
\label{highscale}
\ee

From Eqs.~(\ref{output}) and~\eqref{deltam} 
we can write the matching conditions as
\be
m^2_{LE}\simeq m^2(\QQm)+\Delta m^2(\QQm)=\frac{1}{\tan^2\beta-1}\widetilde m_1^2-\frac{\tan^2\beta}{\tan^2\beta-1}\widetilde m_2^2
\label{EoM}
\ee
where $\Delta m^2$ is given by Eq.~(\ref{Deltam2}), and the subleading
radiative contribution $2m^2\Delta Z_h$ from the wave function
renormalization is neglected. For $\tan\beta\gg 1$, $\mu = \mathcal
O(100\,\rm{GeV})$ and small $\Delta m^2$, the requirement $m^2(\QQm)=
\mathcal O(100\,{\rm GeV})^2$ implies $\left|m_{H_2}^2\right|=
\mathcal O(100\,$GeV)$^2$, where $m_{H_{1,2}}^2$ are defined in
Eq.~(\ref{algunasdef}). For heavy supersymmetric spectrum this size of
$m_{H_2}^2$ can be generated in the neighborhood of the FP
solution~\cite{Delgado:2014vha,Delgado:2013gza} which is radiatively
led to $m^2_{H_2}=0$. However in the FP parameter region there is no
general reason why $\Delta m^2$ should be small. In fact the presence
of $\Delta m^2$ can distort the standard FP solution usually
considered in the literature. In this section we numerically analyze
the parameter space of these modified FP solutions~\footnote{We remind
  that, in the absence of $\Delta m^2$, the FP solution is scale
  invariant with respect to a common multiplicative factor on the
  boundary conditions (at the scale $\mathcal M$) of the
  supersymmetric masses. This scale invariance is broken by $\Delta
  m^2$ which contains logarithms of the supersymmetry breaking masses
  over $\QQm$. Still, as radiative corrections are small as compared
  to the tree level values, the scale invariance of the FP solutions
  is approximatively preserved}.

As it was mentioned in the previous section, it is useful to re-express the supersymmetric parameters of Eq.~(\ref{EoM}) in terms of their values at the high scale $\mathcal M$.  On dimensional grounds we can rewrite them as
\begin{align}
m_X^2(\mathcal Q_m)&=m_X^2+ \eta_Q^X[\mathcal Q_m,\mathcal M](m_Q^2+m_U^2+m_2^2)+
\sum_{a,b}\eta_{ab}^X[\mathcal Q_m,\mathcal M]M_a M_b\nonumber\\
&+\sum_{a}\eta_{aA}^X[\mathcal Q_m,\mathcal M]M_a A_0+\eta_{A}^X[\mathcal Q_m,\mathcal M]A_0^2\nonumber~,\\
A_t(\mathcal Q_m)&=A_0+f_A[\mathcal Q_m,\mathcal M]A_0+\sum_a f_a[\mathcal Q_m,\mathcal M]M_a
\label{carlos}~,
\end{align}
where $A_0$, $m^2_X$ and $M_a$ (with $X=H_i, Q_L, U_R, D_R$ and $a=1,2,3$) are respectively the stop tri-linear mixing parameter, the sfermion masses and the Majorana gaugino masses at the scale $\mathcal M$.

Concerning the numerical procedure, we assume moderately large values of  $\tan\beta$ (namely $\tan\beta\simeq 10$) which allows to approximate $X_t\simeq A_t$ and to safely neglect all Yukawa couplings except that of the top quark. Moreover we set $A_t$ at the mass scale $ \widetilde
{\mathcal Q}^2\equiv m_Q(\mathcal Q_m)m_U(\mathcal Q_m)$ in such a way
that the SM RG evolution of the quartic coupling
\be
\lambda(\widetilde{\mathcal Q})=\frac{1}{4}(g_2^2(\widetilde{\mathcal Q})+g_Y^2(\widetilde{\mathcal Q}))\cos^2 2\beta+\frac{3}{8\pi}y_t^4(\widetilde{\mathcal Q})X_t^2\left(1-\frac{X_t^2}{12}\right)
\ee
from $\widetilde{\mathcal Q}$ to $\mathcal Q_{EW}$ reproduces the
Higgs mass observation, namely $\lambda(\mathcal Q_{EW}) \approx
(m_h/v)^2/2$. For the functions $\eta^X[\mathcal Q_m,\mathcal M]$ and
$f[\mathcal Q_m,\mathcal M]$, which were obtained semi-analytically for
$\mathcal Q_m=2\,$TeV and $\mu\sim 100\,$GeV in
Ref.~\cite{Delgado:2014vha}, we use some simple generalized formulas
where the effect of possible heavy Higgsinos is
incorporated~\footnote{For the case of heavy Higgsinos we determine
  the one-loop RG evolution of the MSSM parameters by neglecting the
  scale dependence of $\mu$. This is justified by the fact that the
  variation of $\mu$ between $\QQm$ and $\mathcal M$  is of the
  order of 1\%.}. Finally, as fundamental description of the
supersymmetric parameters, we consider the relations in Eq.~\eqref{highscale}.
\begin{figure}[htb]
\vskip .5cm 
 \begin{center}
 \includegraphics[width=5.2cm]{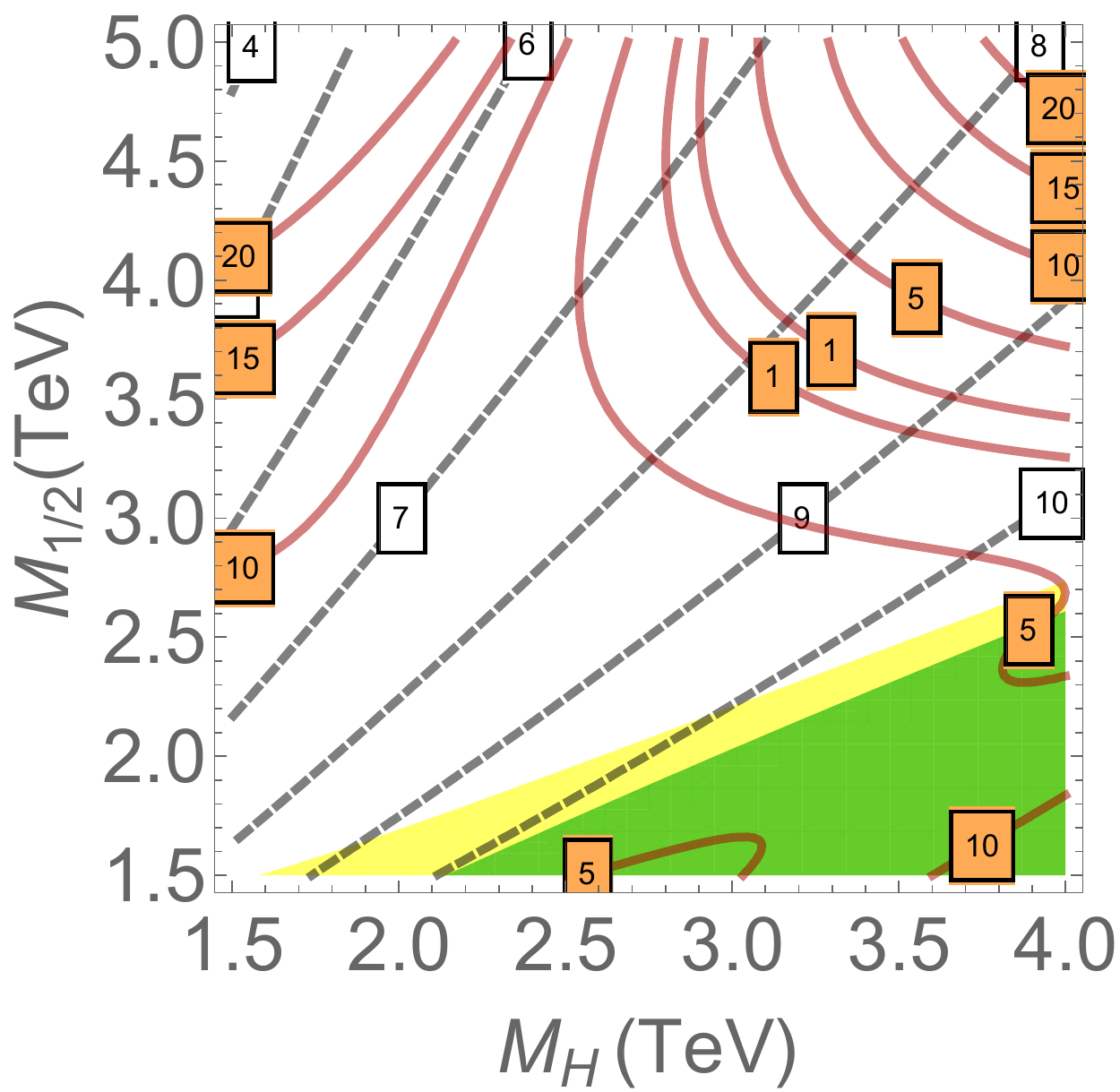}     
 \includegraphics[width=5.2cm]{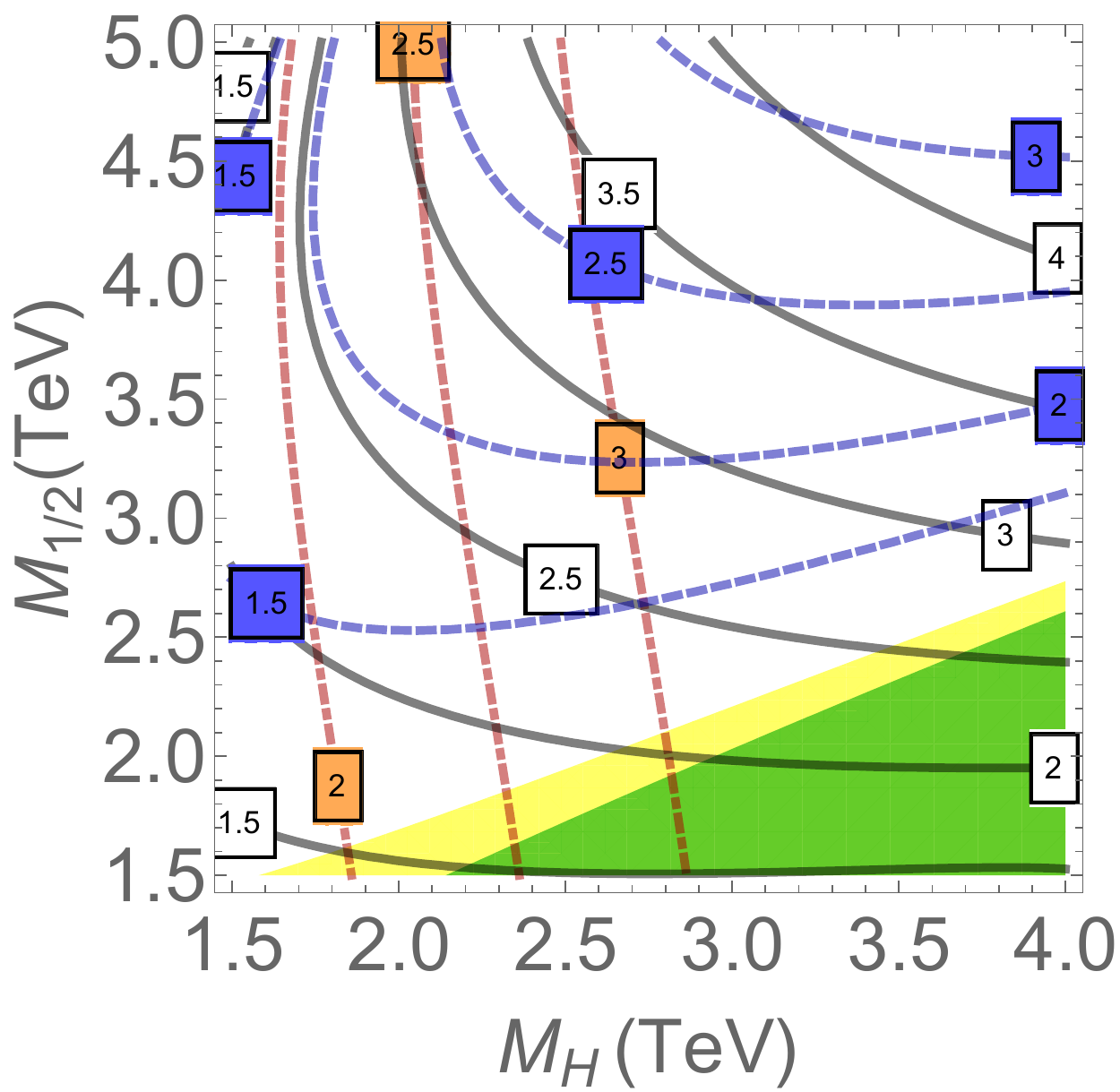}  
    \includegraphics[width=5.2cm]{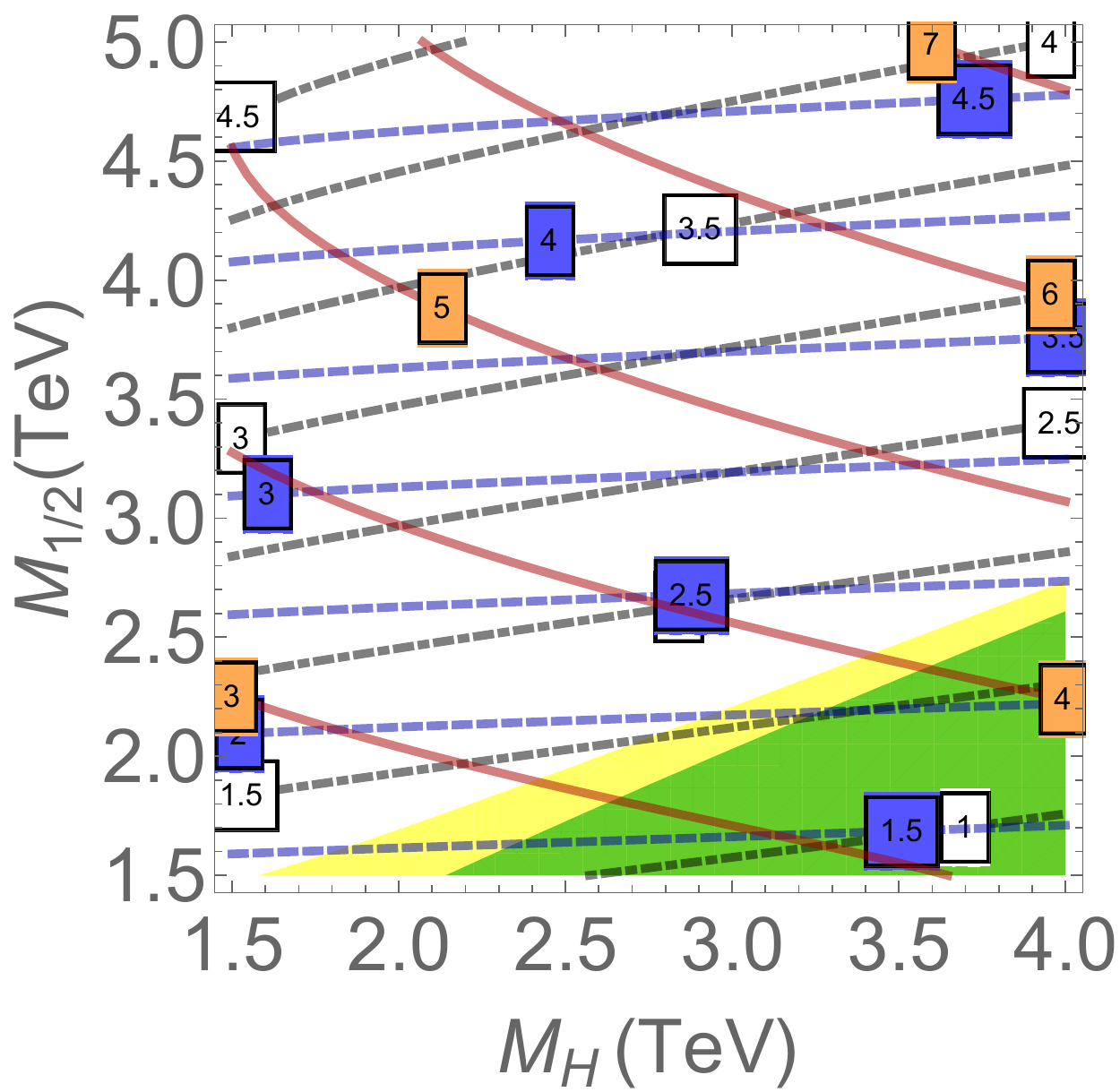}  
\end{center}
\caption{\it\baselineskip=15 pt Left panel: Contour lines of
  $\log_{10}\mathcal M/GeV$ (white labels) and stability parameter
  $\mathcal S$ (orange labels) for $M_0=0$, and light Higgsinos,
  $\mu=100$ GeV. The (yellow) external shadowed region corresponds to
  $m_U(\mathcal Q_m)<750$ GeV and the (green) internal shadowed region
  corresponds to $m_U^2(\mathcal Q_m)<0$. Middle panel: Corresponding
  contour lines of constant $m_Q(\mathcal Q_m)$ (solid black),
  $m_U(\mathcal Q_m)$ (dashed blue) and $m_H(\mathcal Q_m)$ (dash
  dotted red).  Right panel: Corresponding contour lines of constant
  $M_3(\mathcal Q_m)$ (solid red), $M_2(\mathcal Q_m)$ (dashed blue)
  and $M_1(\mathcal Q_m)$ (dash dotted black). Labels in the middle
  and right panels are in TeV units.}
\label{0mulight}
\end{figure}
\begin{figure}[htb]
\vskip .5cm 
 \begin{center}
 \includegraphics[width=5.2cm]{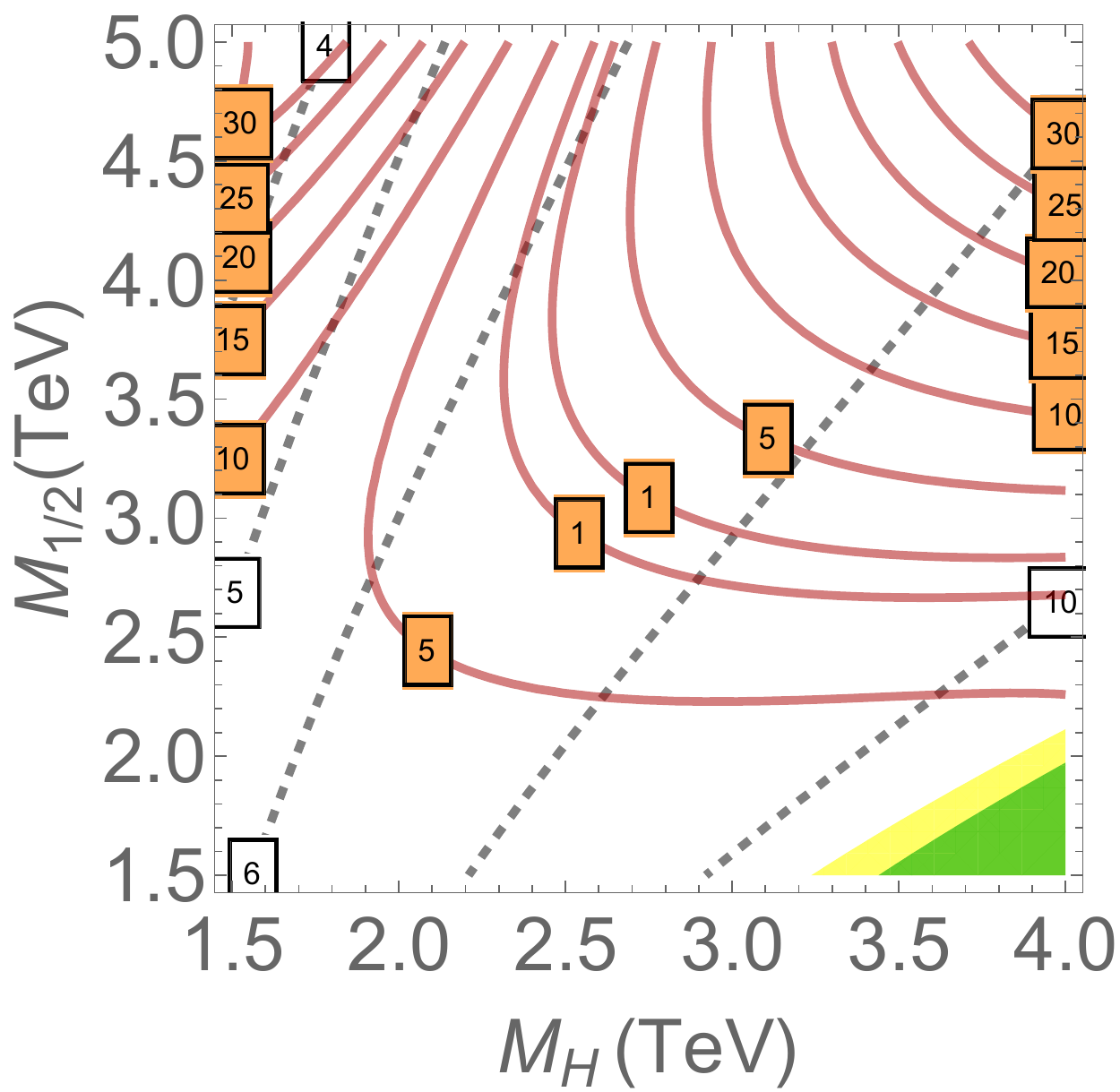}     
 \includegraphics[width=5.2cm]{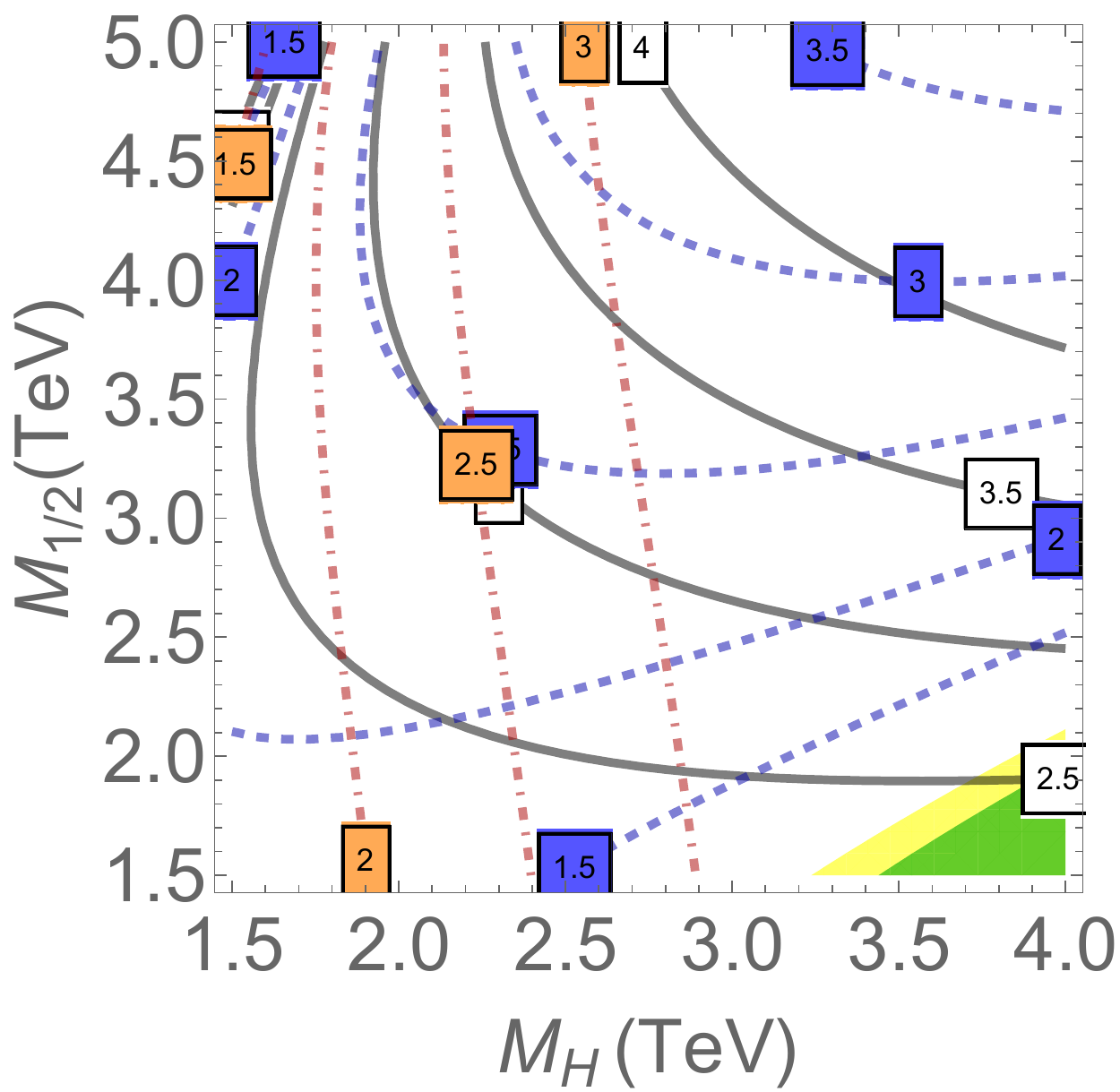}  
    \includegraphics[width=5.2cm]{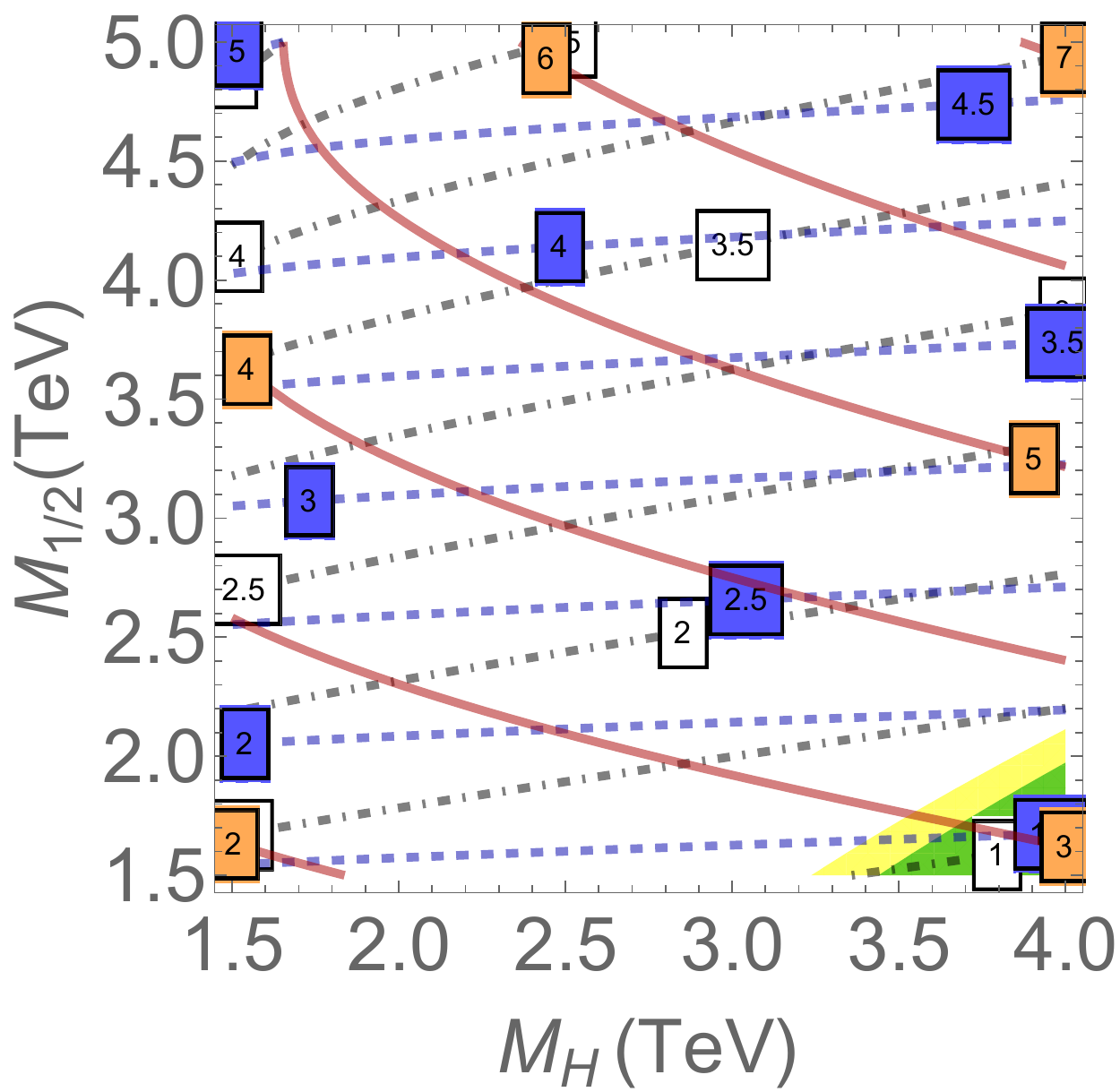}  
\end{center}
\caption{\it The same as in Fig.~\ref{0mulight} but for $M_0=2$ TeV. }
\label{2mulight}
\end{figure}

\begin{figure}[htb]
\vskip .5cm 
\begin{center}
\includegraphics[width=5.2cm]{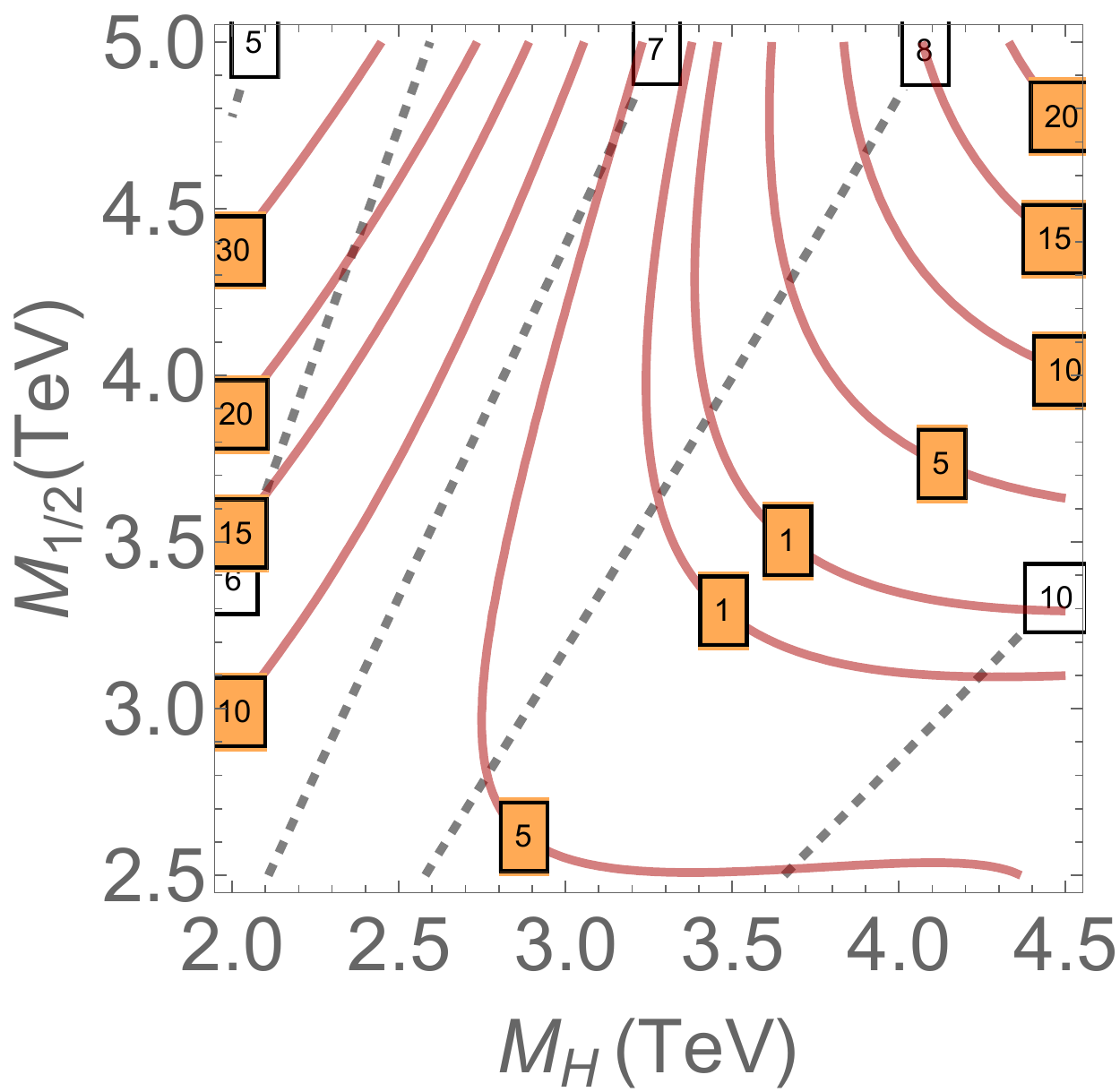}     
\includegraphics[width=5.2cm]{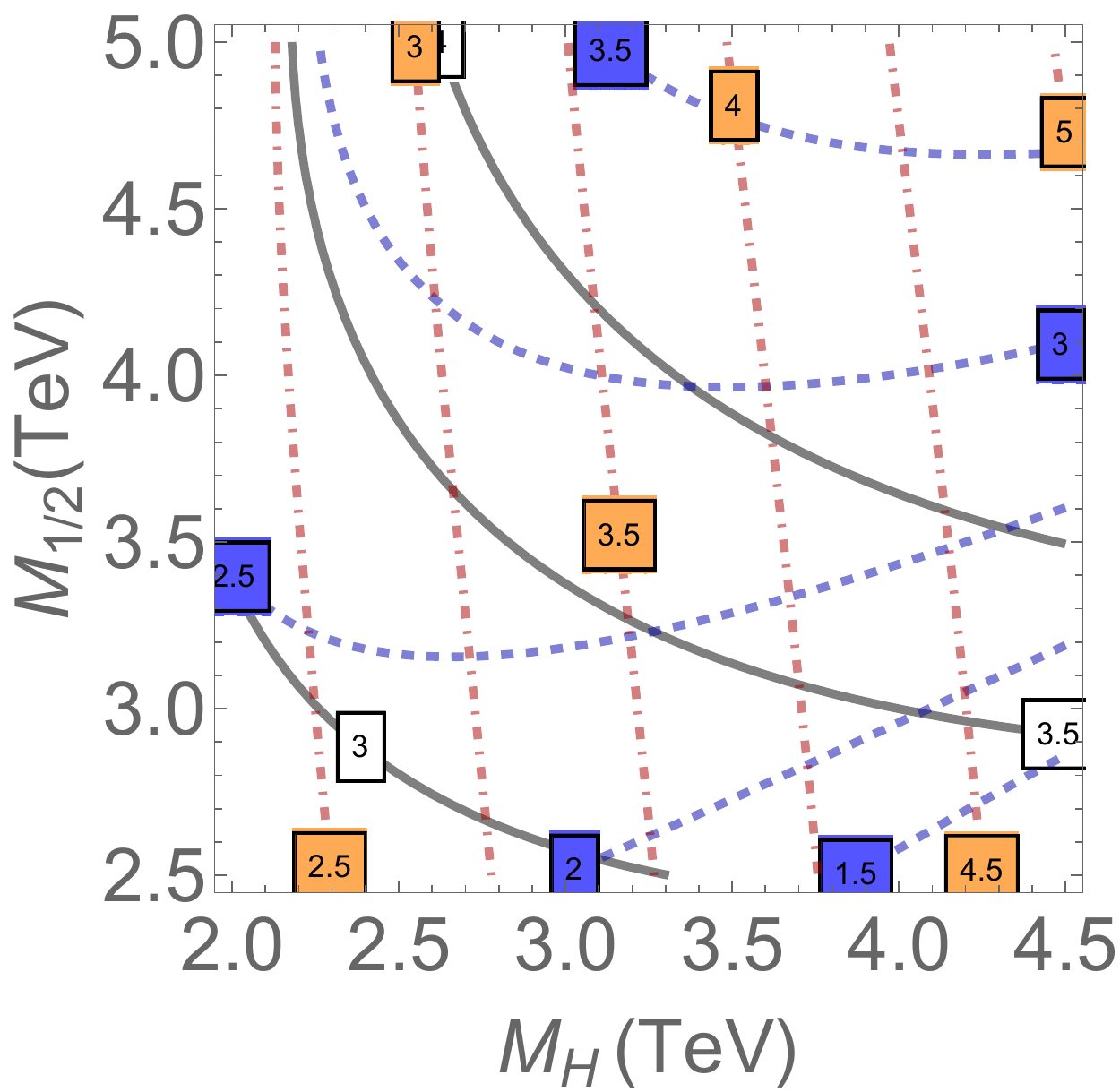}  
\includegraphics[width=5.2cm]{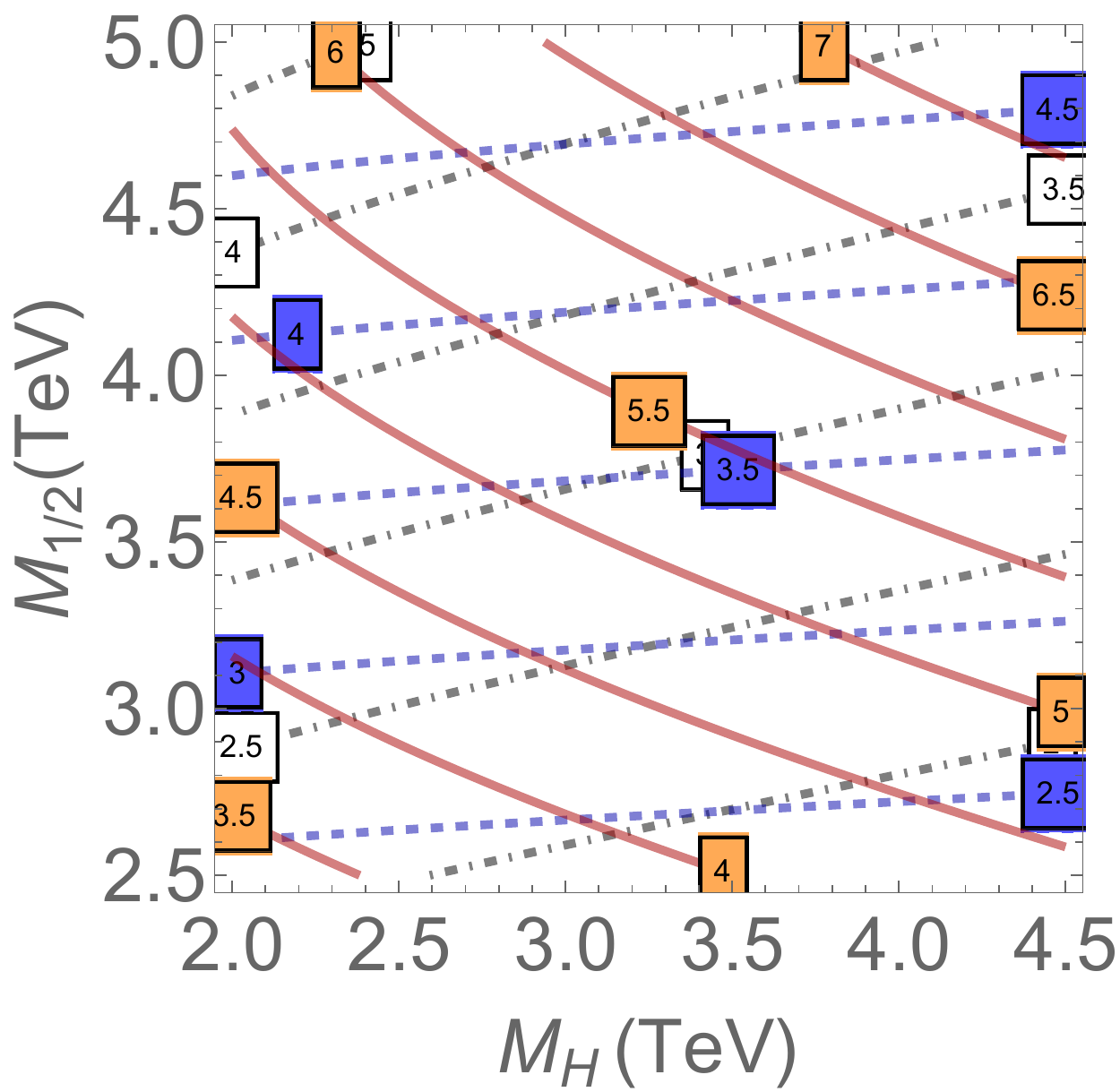}  
\end{center}
\caption{\it The same as in Fig.~\ref{2mulight} but for $\mu=2$ TeV. }
\label{2mu2}
\end{figure}

\begin{figure}[htb]
\vskip .5cm 
 \begin{center}
\includegraphics[width=5.2cm]{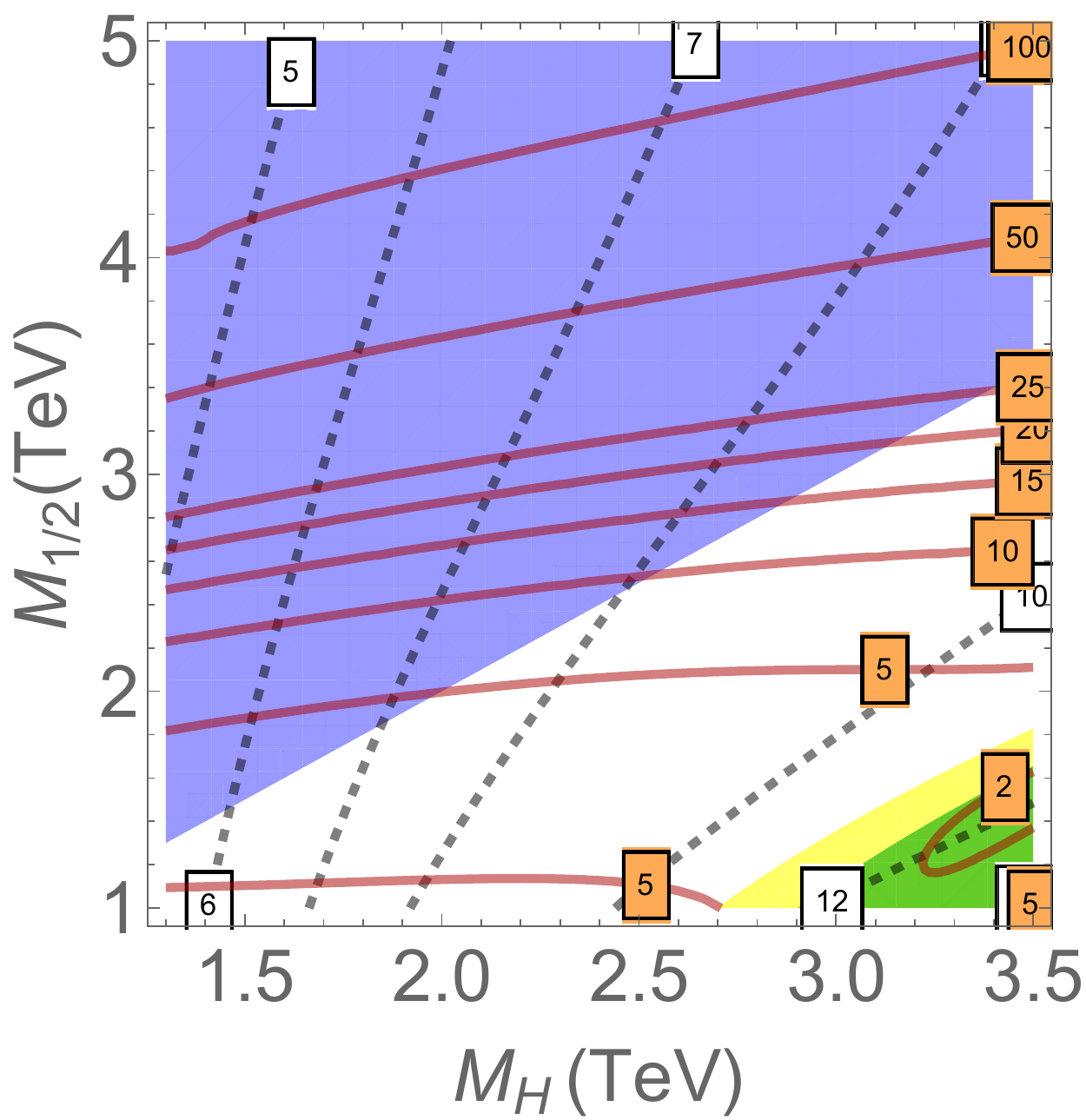}     
\includegraphics[width=5.2cm]{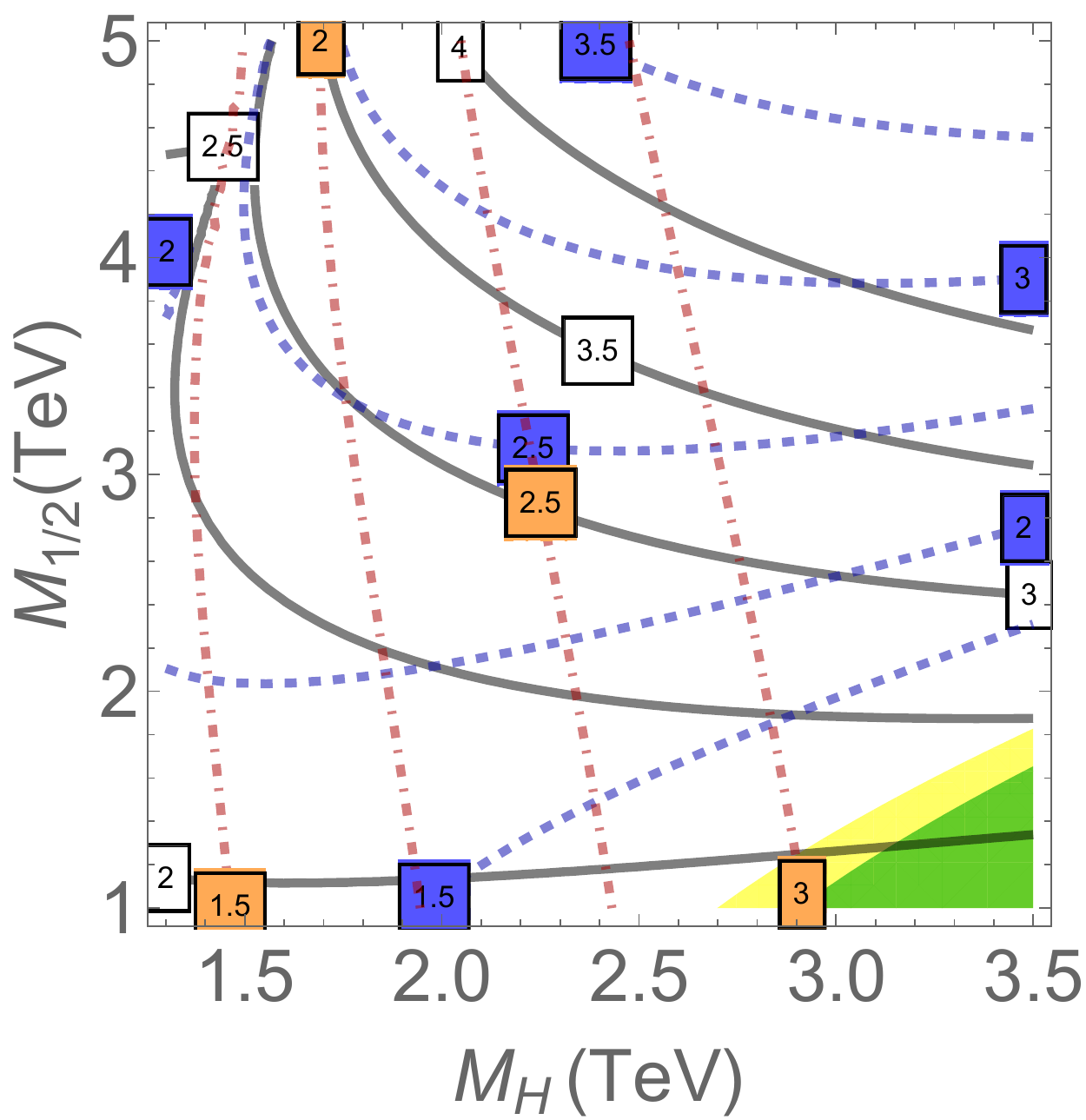}  
   \includegraphics[width=5.2cm]{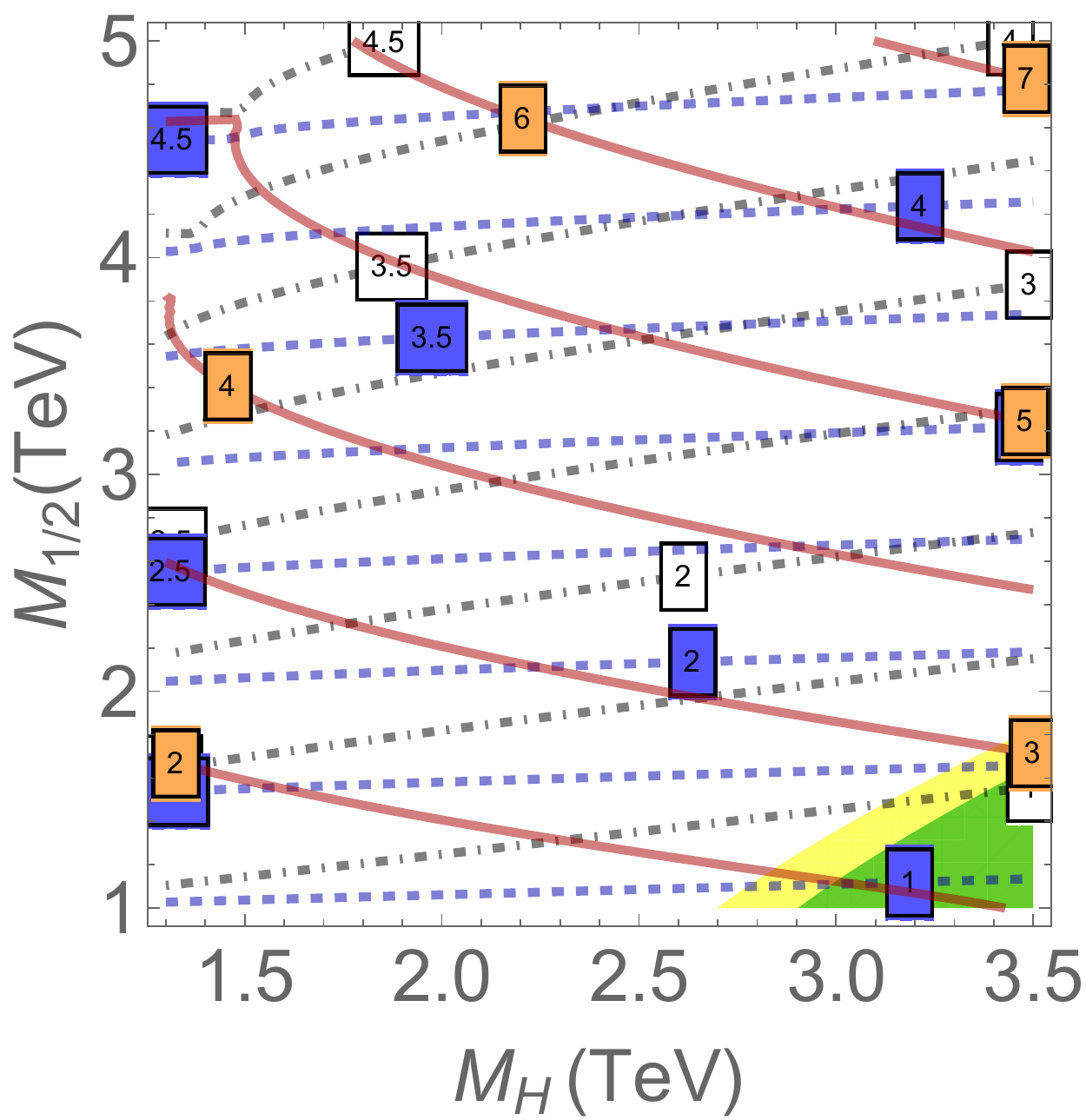}  
\end{center}
\caption{\it \baselineskip=15 pt The same as in Fig.~\ref{2mulight} but for $\mu=M_{1/2}$. In the (magenta) upper left shadowed region the soft breaking terms $m_{H_U}^2=m_{H_D}^2<0$ are tachyonic at the scale $\mathcal M$, although $m_1^2=m_2^2\equiv M_H^2>0$.}
\label{2mum12}
\end{figure}

Figs.~\ref{0mulight}, \ref{2mulight}, \ref{2mu2} and \ref{2mum12}
display the values of fundamental parameters and their corresponding
mass spectra leading to $m^2_{LE}\sim (100\, {\rm GeV})^2$.  In the
left panels of the figures we plot contour (black dashed) lines for
constant values of $\log_{10}(\mathcal M/\textrm{GeV})$ such that
Eq.~\eqref{EoM} at tree level is fulfilled, and contour (solid red)
lines for the stability parameter $\mathcal S$ of the one-loop
radiative corrections (cf.~Eqs.~\eqref{Deltam2} and
\eqref{stability}). Similarly, the contour lines of $m_Q(\mathcal
Q_m)$ (solid black), $m_U(\mathcal Q_m)$ (dashed blue) and
$m_H(\mathcal Q_m)$ (dash-dotted red) are shown in the middle panels,
whereas the contour lines of $M_3(\mathcal Q_m)$ (solid red),
$M_2(\mathcal Q_m)$ (dashed blue) and $M_1(\mathcal Q_m)$ (dash-dotted
black) are depicted in the right panels. We remind that the condition
$m^2_{LE}\sim (100\,{\rm GeV})^2$ arises with no tuning between $m^2$
and $\Delta m^2$ when $\mathcal S=\mathcal O(1)$.


In Figs.~\ref{0mulight} and \ref{2mulight} we consider the case $\mu\sim 100\,$GeV, and $M_0=0$, $M_0=2$ TeV, respectively, and in Figs.~\ref{2mu2} and \ref{2mum12} the case with $M_0=2$ TeV and heavy and superheavy Higgsinos, with $\mu=2$ TeV and $\mu=M_{1/2}$, respectively. The region with $m_U(\mathcal Q_m)<750$ GeV [$m_U^2(\mathcal Q_m)<0$] corresponds to the yellow [green] shadowed area. In fact the yellow band corresponds to the FP for the light stop
scenario~\cite{Delgado:2014kqa}.  

As we can see from the general expression of radiative corrections, Eq.~(\ref{Deltam2}), gauge coupling terms should tend to compensate top Yukawa coupling terms when $M_{1/2},\, M_H>M_0$~\footnote{This kind of spectra, where the boundary conditions for
gauginos are heavier than sfermions, can be found e.g.~in minimal gauge
mediation (although the considered boundary conditions in our example of Eq.~(\ref{highscale}) do not match those of minimal gauge mediation) with a largish number of messengers or in some extra dimensional mechanisms of supersymmetry breaking as gaugino
mediation~\cite{Giudice:1998bp,Antoniadis:1998sd}.}. This is the case for the models analyzed in Figs.~\ref{0mulight}-\ref{2mu2} where the stability condition is satisfied for $M_{1/2},\, M_H\gtrsim 3$ TeV. 
Finally this trend is broken for superheavy Higgsinos, as the $\mu=M_{1/2}$ case shown in Fig.~\ref{2mum12}. Here for very heavy gauginos, $M_{1/2}\gg 1$ TeV, the value of $M_H^2=m_{H_{u,d}}^2+\mu^2$ becomes very large and the condition $\mathcal S=1$ is overshot. Therefore the stability condition requires lighter gauginos than scalars as we can see in Fig.~\ref{2mum12}.

%
%


\section{\sc Outlook and conclusions}
\label{conclusion}

In view of the growing experimental evidence against the existence of
sub-TeV new physics, and having in mind the naturalness problem of
electroweak interactions, it is interesting to study the matching of
the SM with possible UV completions solving the grand hierarchy
problem. In general we expect that the presence of heavy mass
states, coupled to the Higgs sector in the UV theory, would contribute
at the matching scale $\mathcal Q_m$ as finite threshold corrections
to the SM Higgs mass term: these finite contributions trigger
the hierarchy problem and destabilize the electroweak minimum. Determining these corrections is then of
utmost importance to quantify the stability/naturalness problem and at least to
find the parameter conditions that the underlying (final) theory at
the unification, or Planck, scale should eventually provide to solve
it.

In this paper we have pursued this task for the simplest
perturbative UV completion of the SM solving the hierarchy problem:
the MSSM. Although we have focused on this model, the qualitative
features presented in this paper are expected to hold in any
perturbative UV theory aiming to solve the grand hierarchy
problem. The key ingredient of our analysis is the one-loop effective
potential improved by the renormalization group equations, and the
renormalization-scale invariance of such a potential.  As a battleground
we have considered the Landau gauge and dimensional regularization in
the $\overline{MS}$ renormalization scheme (at this level of the
calculation, and for scalar and fermion fields integrated out, 
equivalent to the $\overline{DR}$). With these choices
we have obtained the one-loop matching between the SM and MSSM Higgs
sectors at the multi-TeV matching scale $\mathcal Q_m$, where the large leading logarithms between the high (unification) scale $\mathcal M$ and $\mathcal Q_m$ have been resummed. 

Motivated by hints from experimental data, the matching
has been performed assuming a large mass hierarchy between SM and
non-SM fields, which allows to work in the unbroken electroweak symmetry phase.
After having integrated the heavy fields out, the final one-loop
identification between the SM mass term, $m^2_{SM}$, and the MSSM mass
terms, $m_I^2$ (with $I=1,2,3$), is given by Eq.~(\ref{deltam}) which
can be written as
\be
m_{SM}^2=\left[-m_1^2 c^2_\beta-m_2^2 s^2_\beta+2 m_3^2s_\beta c_\beta\right](1+2\Delta Z_h)+\Delta m^2
\label{ultima}~,
\ee
where the radiative contributions $\Delta Z_h$ and $\Delta m^2$
  are respectively coming from the matching between the wave functions
  and quadratic interactions of the SM and the MSSM lightest CP-even
Higgs.  The main features of the matching are as follows:
\begin{itemize}
\item We have integrated out only heavy states from the MSSM in the
  effective potential. We have \textit{not} integrated out the heavy modes
  from the light (SM) states, as in the Wilsonian action, which would
  have created a cutoff in the low energy theory. In this way we can
  still integrate momenta in the low energy theory up to infinity and
  then keep on using dimensional regularization for them.
\item Using the scale independence of the effective potential the
  matching scale $\mathcal Q_m$ is completely arbitrary in the
 considered one-loop approximation. If all heavy masses are of similar order of
  magnitude (as in the high-scale supersymmetry scenario considered
  here), $\mathcal Q_m$ can be arbitrarily fixed at some intermediate
  value around them to avoid large logarithms and the breaking of
  perturbation theory at higher-loop orders.  If very different heavy scales
  are present (as e.g.~in split
  supersymmetry~\cite{ArkaniHamed:2004yi}) then different matching
  processes should be subsequently applied
  (cf.~e.g.~Ref.~\cite{Carena:2008rt}).
\item The size of the squared mass parameter of the lightest-Higgs in
  the high energy theory, $m^2=-m_1^2 c^2_\beta-m_2^2 s^2_\beta+2
  m_3^2s_\beta c_\beta$, is not physically meaningful as it is
  strongly scale dependent.  Its dependence is mostly compensated by
  the one of $2\Delta Z_h$ and $\Delta m^2$, so that the final
  one-loop scale dependence of the right hand side~of
  Eq.~(\ref{ultima}) is as weak as that of the SM mass parameter (in
  particular the dependence including large masses appears only at two
  loops). The natural scale of the SM electroweak sector should hence
  be deduced considering the sum of $m^2$ and $\Delta m^2$ (the
  $\Delta Z_h$ effect is in general subleading), possibly at the scale
  $\QQm$ to avoid perturbative problems. In particular, once the
  magnitude of $m^2(\QQm)$ is enforced to be of the order of the
  electroweak scale, the stability of the electroweak breaking
  conditions is guaranteed by $\mathcal S\equiv |\Delta
  m^2(\QQm)/m^2_{LE}(\QQm) |\lesssim 1$.
\end{itemize}

We now present a short list of results obtained in the present paper:
\begin{itemize}
\item When the non-SM fields are heavy and degenerate, the expression
  for $\Delta m^2$ at the matching scale $\mathcal Q_m$ reproduces
  the result obtained by Veltman in the SM using dimensional
  regularization and extracting the ``quadratic" divergence as the
  residue of the pole in $d=2$ dimensions. Veltman then interpreted this result as
  the coefficient of the cutoff $\Lambda^2$, while we can express it as
  the coefficient of the common supersymmetric mass squared. Our result is
  thus consistent with Veltman's and puts solid grounds in the
  understanding of the SM as an effective theory below the MSSM. As
  it is well known, the vanishing of the Veltman coefficient can only be
  achieved at (super)Planckian scales.
\item When the MSSM fields are heavy, and of the same order of magnitude but not fully degenerate, the stability of the electroweak minimum provides a generalization of Veltman's result, which
  amounts to introducing different cutoffs for the different SM
  sectors, plus a modification that is negligible only for a
 strictly non-hierarchical heavy spectrum. 
 In
    particular by assuming the masses of the heavy Higgses and/or gauginos to be
    larger than those of sfermions, one can
  easily achieve the vanishing of $\Delta m^2$ at sub-Planckian scales.

\item We have analyzed several cases where the MSSM mass parameters
  are generated at high scale, and transmitted to the observable sector by means of some mediators between the
  hidden and observable sectors, and we have determined the parameter
  regions where $m^2(\QQm)\sim (100\,\rm{GeV})^2$. However also the requirement $\Delta m^2(\mathcal Q_m)\lesssim
    \mathcal O( 100\,\textrm{GeV})^2$ should be imposed in order to
    achieve satisfactory electroweak breaking conditions. The
  parameter regions where also this latter requirement is realized
  turn out to provide a sort of generalized focus point conditions
  which include threshold effects.
\end{itemize}
Let us finally make some considerations about the physical
  significance of the (stability) regions where both $m^2_{LE}(\mathcal
  Q_{EW})=\mathcal O(100\, \textrm{GeV})^2$ and $\mathcal S \lesssim 1$ conditions
  are satisfied:
\begin{itemize}
\item As we have already mentioned, these regions are obtained by
  integrating out the heavy fields at one-loop. The procedure is
  performed at a renormalization scale nearby the energy scale of the
  heavy masses. The result is thus not jeopardized by large logarithms
  and it has no relevant one-loop scale dependence.
\item
The stability regions do depend on the threshold contributions
    we obtain from the effective potential in a particular (Landau) gauge. 
    Therefore, in general, we could expect a gauge
    dependence in our results. However as we have performed the
  matching in the unbroken electroweak symmetry phase and upon
  integrating out only scalars and fermions (neither gauge nor
  Goldstone bosons), it turns out that our computation of the
    thresholds is \textit{gauge independent} within the considered
  approximations.
\item The effective potential depends on the particular
  renormalization scheme, and we have worked it out in the
  $\overline{MS}$ scheme which amounts to subtracting (to define the
  counter-term) the infinite term proportional to
  $\frac{2}{\epsilon}-\gamma_E+\log(4\pi)-\delta$ with $\delta=0$.  In
  the $\overline{MS}$ scheme the finite term in the effective
  potential contributions coming from heavy fermions and scalars is
  proportional to the constant $\mathcal C=3/2$
  [cf.~Eq.~\eqref{CW}]. Subtracting a different infinite counter term
  (with $\delta\ne 0$, as e.g.~in the $MS$ renormalization scheme or a
  variant thereof) would then lead to a shift in the constant $C$ as
  $C\to C+\delta$. Consequently for the degenerate and almost
  degenerate cases, the stability condition $\mathcal S= 0$ leading
  respectively to the exact and generalized Veltman conditions
  (cf.~sections~\ref{degener} and \ref{nondegener} with $r_{1/2}\simeq
  r_H \simeq 1$) would receive additional contributions. Therefore,
  the Veltman-like conditions we obtain from the MSSM arise
  \textit{only} in the $\overline{DR}$ scheme and are relevant for the
  matching with the SM effective theory in the $\overline{MS}$
  renormalization scheme.
\end{itemize}

To conclude, given a scheme of supersymmetry breaking involving soft
  terms above the TeV scale, the threshold corrections to the Higgs
  mass parameters play a determinant role in the stability of the
  electroweak breaking conditions and the masses of the Higgs
  fields. In the present paper we have determined these threshold
  corrections in some MSSM scenarios with no relevant hierarchy in the
  heavy mass spectrum, and looked for soft-parameter relations that
  could alleviate the little hierarchy problem. Of course it is
clear that a similar analysis can also be performed for any perturbative
theory that UV completes the SM. Moreover an equivalent analysis can
be done even if the low energy theory is itself some extension of
the SM, as one in which there is an aligned extended Higgs
sector~\cite{Gunion:2002zf} giving rise to e.g.~a two Higgs doublet
model. If the UV completion of the SM is not perturbative, as in the
case of a composite Higgs, the calculation cannot rely on perturbation
theory and different methods to evaluate threshold effects should be
used~\cite{Tavares:2013dga}. In general, whatever the final UV
completion of the SM is, we expect it could provide an answer to the
question on why our electroweak vacuum is stable and insensitive to
high scale physics.

\section*{Acknowledgments}
\noindent
We are grateful to Pietro Slavich for reading the
 first version of the manuscript and providing useful comments. The work of I.M.~is
supported by the research grant ``Theoretical Astroparticle Physics"
number 2012CPPYP7 under the program PRIN 2012 funded by the Ministero
dell'Istruzione, Universit\`a e della Ricerca (MIUR).  The work of
G.N.~is supported by the German Science Foundation (DFG) within the
Collaborative Research Center (SFB) 676 ``Particles, Strings and the
Early Universe''.  The work of M.Q.~is partly supported by the Spanish
Consolider-Ingenio 2010 Programme CPAN (CSD2007-00042), by
CICYT-FEDER-FPA2011-25948, by the Severo Ochoa excellence program of
MINECO under the grant SO-2012-0234, by Secretaria d'Universitats i
Recerca del Departament d'Economia i Coneixement de la Generalitat de
Catalunya under Grant 2014 SGR 1450, and partly done at IFT and
ICTP-SAIFR under CNPq grant 405559/2013-5.  We also acknowledge
support from CERN, where this work was partly done.

\appendix
\section{\sc Radiative corrections to mass parameters}
\label{appendixA}
\noindent
In this appendix we will present a detailed calculation of the radiative contributions to the mass parameters $\Delta m_1^2$, $\Delta m_2^2$ and $\Delta m_3^2$ in the MSSM arising from the different sectors of sfermions, charginos, neutralinos and the Higgs scalar
sector.

\subsection{\sc Squarks and sleptons}
\noindent
The mass squared matrices for top and bottom squarks and for tau-sneutrinos and staus can be written as
\begin{equation}
M^2_{\tilde t} = \left( \begin{matrix} m_Q^2 +Y_t^2 h_2^2 + \Pi_{\tilde u_L} & 
Y_t (A_t h_2 -\mu h_1 ) \\ 
Y_t (A_t h_2 -\mu h_1 ) 
& m_U^2 +Y_t^2 h_2^2  +  \Pi_{\tilde u_R}  \end{matrix} \right)
\end{equation}
\begin{equation}
M^2_{\tilde b} = \left( \begin{matrix} m_Q^2 +Y_b^2 h_1^2 + \Pi_{\tilde d_L} & 
Y_b (A_b h_1 -\mu h_2 ) \\ 
Y_b (A_b h_1-\mu h_2 ) 
& m_D^2 +Y_b^2 h_1^2 + \Pi_{\tilde d_R}  \end{matrix} \right)
\end{equation}
\begin{equation}
M^2_{\tilde\nu_\tau} = \left( \begin{matrix} m_L^2 + \Pi_{\tilde \nu_L} & 
0 \\ 
0 
& m_N^2   \end{matrix} \right)
\end{equation}
\begin{equation}
M^2_{\tilde \tau} = \left( \begin{matrix} m_L^2 +Y_\tau^2 h_1^2 + \Pi_{\tilde e_L} & 
Y_\tau (A_\tau h_1 -\mu h_2 ) \\ 
Y_\tau (A_\tau h_1-\mu h_2 ) 
& m_E^2 +Y_\tau^2 h_1^2+  \Pi_{\tilde e_R}  \end{matrix} \right)
\end{equation}
where
\be
\Pi_{\tilde f}=\frac{1}{2}\left[ T_{3_{\tilde f}} \left(g_Y^2+g_2^2\right)-Q_{\tilde f} g_Y^2\right](h_1^2-h_2^2)\,\,.
\ee
Using the expressions in Eq.~\eqref{Dmsqs} it is straightforward to find the
corresponding contributions to $\Delta m^2_{1,2,3}$ as
\bea
\Delta m_1^2&=&\frac{2}{32\pi^2}\left\{3 Y_b^2[G(m_Q^2)+G(m_D^2)]+Y_\tau^2[G(m_L^2)+G(m_E^2)]+\frac{3 Y_t^2\mu^2}{m_Q^2-m_U^2}[G(m_Q^2)-G(m_U^2)]\right.\nonumber\\
&+&\frac{3 Y_b^2A_b^2}{m_Q^2-m_D^2}[G(m_Q^2)-G(m_D^2)]+\frac{Y_\tau^2A_\tau^2}{m_L^2-m_E^2}[G(m_L^2)-G(m_E^2)]\nonumber\\
&-&\left.\frac{g_Y^2}{2}[G(m_Q^2)-2G(m_U^2)+G(m_D^2)-G(m_L^2)+G(m_E^2)]\right\}\nonumber~,\\
\Delta m_2^2&=&\frac{2}{32\pi^2}\left\{3 Y_t^2[G(m_Q^2)+G(m_U^2)]+\frac{3 Y_t^2 A_t^2}{m_Q^2-m_U^2}[G(m_Q^2)-G(m_U^2)]\right.\nonumber\\
&+&\frac{3 Y_b^2 \mu^2}{m_Q^2-m_D^2}[G(m_Q^2)-G(m_D^2)]+\frac{Y_\tau^2\mu^2}{m_L^2-m_E^2}[G(m_L^2)-G(m_E^2)]\nonumber\\
&+&\left.\frac{g_Y^2}{2}[G(m_Q^2)-2G(m_U^2)+G(m_D^2)-G(m_L^2)+G(m_E^2)]\right\}\nonumber~,\\
\Delta m_3^2&=&\frac{2}{32\pi^2}\left\{\frac{3Y_t^2\mu A_t}{m_Q^2-m_U^2}[G(m_Q^2)-G(m_U^2)]+ \frac{3Y_b^2\mu A_b}{m_Q^2-m_D^2}[G(m_Q^2)-G(m_D^2)]\right.\nonumber \\
&+&\left. \frac{Y_\tau^2\mu A_\tau}{m_L^2-m_E^2}[G(m_L^2)-G(m_E^2)]\right\}\,\,, \nonumber
\eea
with
\be
 G(m_X^2)\equiv m_X^2\left[ \log\frac{m_X^2}{\mathcal Q_m^2}-1-\delta  \right] ~.
\label{Ggeneric}
\ee
Here the function $G$ has been generalized to the subtraction
  schemes having $\mathcal C = 3/2 + \delta$ in the scalar
  contributions of Eq.~\eqref{CW}. Eq.~\eqref{Ggeneric} recovers
  Eq.~\eqref{Gmsbar} for the $\overline{MS}$ (or $\overline{DR}$) subtraction scheme where $\delta=0$.

If some of the masses are degenerate the following limit turns
  useful:
\be
\lim_{y\to x} \frac{G(x^2)-G(y^2)}{x^2-y^2} = \log \frac{x^2}{\QQm^2} -\delta\ .
\label{primera}
\ee

\subsection{\sc Charginos}
\noindent
The squared mass matrix for the charginos can be written as 
\begin{equation}
M^2_{ch} = \left( \begin{matrix} M_2^2+g_2^2 h_2^2 & 
g_2 (M_2h_1+\mu h_2)  \\ 
g_2 (M_2h_1+\mu^*h_2) 
&  \mu^2+g_2^2 h_1^2 \end{matrix} \right)\,\,.
\end{equation}
We can then compute the corresponding contributions to $\Delta m_{1,2,3}^2$ as
\be
\Delta m_1^2=\Delta m_2^2=-\frac{4}{32\pi^2}\frac{g_2^2}{M_2^2-\mu^2}\,[M_2^2G(M_2^2)-\mu^2G(\mu^2)]~,
\ee
\be
\Delta m_3^2=\frac{4}{32\pi^2}\frac{g_2^2\mu M_2}{M_2^2-\mu^2}\,[G(M_2^2)-G(\mu^2)]~ ,
\ee
with $G$ defined in Eq.~\eqref{Ggeneric}. In case of degenerate
  spectra the following limit is useful:
\be
\lim_{y\to x} \frac{x^2 G(x^2)-y^2 G(y^2)}{x^2-y^2} = x^2 \left(-1-2\delta +2\log \frac{x^2}{\QQm^2}\right)
\label{segunda}
\ee
with $\delta=0$ in the $\overline{MS}$ or $\overline{DR}$ schemes.

\subsection{\sc Neutralinos}
\noindent
The squared mass matrix for the neutralinos can be written as 
\begin{equation}
M^2_{ne} = \left( \begin{matrix} 
M_1^2+\frac{g_Y^2}{2}  (h_1^2+h_2^2)  & -\frac{1}{2} g_Y g_2 (h_1^2+h_2^2)  & -\frac{g_Y}{\sqrt{2}}(M_1 h_1+\mu h_2)   & \frac{g_Y}{\sqrt{2}}(M_1 h_2+\mu h_1)\\ 

-\frac{1}{2} g_Y g_2 (h_1^2+h_2^2) & M_2^2+\frac{g_2^2}{2}  (h_1^2+h_2^2) & \frac{g_2}{\sqrt{2}}(M_2 h_1+\mu h_2)   & -\frac{g_2}{\sqrt{2}}(M_2 h_2+\mu h_1) \\ 

-\frac{g_ 1}{\sqrt{2}}(M_1 h_1+\mu h_2)   & \frac{g_2}{\sqrt{2}}(M_2 h_1+\mu h_2)   & \mu^2+\frac{h_1^2}{2}  (g_Y^2+g_2^2)  & -\frac{1}{2} h_1 h_2  (g_Y^2+g_2^2) \\

\frac{g_Y}{\sqrt{2}}(M_1 h_2+\mu h_1)   & -\frac{g_ 2}{\sqrt{2}}(M_2 h_2+\mu h_1)  & -\frac{1}{2} h_1 h_2  (g_Y^2+g_2^2) & \mu^2+ \frac{h_2^2}{2}  (g_Y^2+g_2^2) 
\end{matrix} \right)\,\,.
\end{equation}
The derivatives of the mass eigenvalues with respect to the backgrounds $\phi_j\equiv(h_1^2,h_2^2,h_1 h_2)$ can be easily computed following the techniques introduced in Ref.~\cite{Ibrahim:2002zk}. The squared mass eigenvalues are given by the solutions of the equation defined by the characteristic polynomial
\be
\det(M^\dagger M-\lambda)\equiv\sum_n c^{(n)}(\phi_j)\lambda^n=0
\label{caracteristico}
\ee
where the coefficients of the characteristic polynomial are functions of $h_1^2$, $h_2^2$ and $h_1 h_2$. Differentiating (\ref{caracteristico}) with respect to $\phi_j$ we obtain the required expressions
\be
\left.\frac{\partial\lambda}{\partial \phi_j}\right|_{h_i=0}=-\left.\frac{\sum_n \frac{\partial c^{(n)}(\phi_j)}{\partial \phi_j}\lambda^n}{\sum_n nc^{(n)}(\phi_j)\lambda^{n-1}}\right|_{h_i=0}
\label{master}
\ee
where on the right hand side~$\lambda$ denotes the squared mass eigenvalues.

The calculation of $\Delta m_{1,2,3}^2$ is now straightforward and gives
\begin{align}
\Delta m_1^2&=
\Delta m_2^2=-\frac{2}{32\pi^2}\left\{\frac{g_Y^2}{M_1^2-\mu^2}[M_1^2G(M_1^2)-\mu^2G(\mu^2)]\right.\nonumber\\
&+\left.\frac{g_2^2}{M_2^2-\mu^2}[M_2^2G(M_2^2)-\mu^2G(\mu^2)]\right\}\nonumber\\
\Delta m_3^2&=\frac{2}{32\pi^2}\left\{\frac{g_Y^2 M_1 \mu}{M_1^2-\mu^2}[G(M_1^2)-G(\mu^2)]+\frac{g_2^2 M_2 \mu}{M_2^2-\mu^2}[G(M_2^2)-G(\mu^2)]\right\}\,\,.
\end{align}
In the case of equal masses we can use the limiting behavior in Eqs.~(\ref{primera}) and (\ref{segunda}).

\subsection{\sc Higgs scalar sector}
\noindent
The general tree level potential for the scalar sector is given by
\begin{align}
V&=m_1^2 |H_1|^2+m_2^2|H_2|^2+m_3^2(H_1\cdot H_2+h.c.)+\frac{g_Z^2}{2}  (|H_2|^2-|H_1|^2)^2\nonumber\\
&+\frac{g_2^2}{2}|H_2^+ H_1^{0*}+H_1^{-*}H_2^0|^2
\end{align}
where $H_1\cdot H_2=H_1^a\varepsilon_{ab}H_2^b$ with $\varepsilon_{12}=-1$.
The squared mass matrices for scalars [in the basis $(\textrm{Re}\, H_1^0,\textrm{Re}\, H_2^0)$], pseudoscalars [in the basis $(\textrm{Im}\, H_1^0,\textrm{Im}\, H_2^0)$] and charged scalars [in the basis $(H_2^+,H_1^{-*})$] are
\begin{equation}
M^2_{S} = \left( \begin{matrix} m_1^2-g_Z^2(h_2^2-3 h_1^2) & -m_3^2-2 g_Z^2 h_1 h_2
 \\ 
-m_3^2-2 g_Z^2 h_1 h_2 
&  m_2^2+g_Z^2(3h_2^2-h_1^2) \end{matrix} \right)
\end{equation}
\begin{equation}
M^2_{P} = \left( \begin{matrix} m_1^2-g_Z^2(h_2^2-h_1^2) & m_3^2
 \\ 
m_3^2 
&  m_2^2+g_Z^2(h_2^2-h_1^2) \end{matrix} \right)
\end{equation}
\begin{equation}
M^2_{C} = \left( \begin{matrix} m_2^2+g_Z^2(h_2^2-h_1^2)+\frac{g_2^2}{2}h_1^2 & m_3^2+\frac{g_ 2^2}{2} h_1 h_2
 \\ 
m_3^2+\frac{g_ 2^2}{2} h_1 h_2 
&  m_1^2-g_Z^2(h_2^2-h_1^2)+\frac{g_2^2}{2}h_2^2\end{matrix} \right)\,\,.
\end{equation}

For each of the previous matrices there are two eigenstates in the unbroken phase ($h_i=0$), $H_{SM}=c_\beta H_1+s_\beta \tilde H_2^* $ with mass eigenvalue $-m^2$, and $H_h=-s_\beta H_1+c_\beta\tilde H_2^*$ with mass eigenvalue $m_H^2=m_1^2+m_2^2+m^2$. The doublet $H_{SM}$ is identified with the SM Higgs doublet, and we then exclude its contribution to $\Delta m^2$. Using Eq.~\eqref{master} one obtains that the heavy Higgs contributions are given by
\begin{align}
\Delta m_1^2&=\frac{1}{32\pi^2}\, \frac{G(m_H^2)}{m_1^2+m_2^2+2m^2}\left[2g_Z^2(3 m_1^2-2m_2^2+m^2)+g_2^2(m_2^2+m^2)
\right]\nonumber\\
\Delta m_2^2&=\frac{1}{32\pi^2}\, \frac{G(m_H^2)}{m_1^2+m_2^2+2m^2}\left[2g_Z^2(3 m_2^2-2m_1^2+m^2)+g_2^2(m_1^2+m^2)
\right]\nonumber\\
\Delta m_3^2&=\frac{1}{32\pi^2}\, \frac{G(m_H^2)}{m_1^2+m_2^2+2m^2}\left[-2g_Z^2-g_2^2
\right]m_3^2~.
\label{scalars1}
\end{align}
Using now the matching conditions \eqref{matching1}, Eqs.~\eqref{scalars1} can be also written as
\begin{align}
\Delta m_1^2&=\frac{1}{32\pi^2}\, G(m_H^2)\left[
-4g_Z^2 \cos 2\beta+2g_Z^2 \sin^2\beta+g_2^2\cos^2\beta
\right]\nonumber\\
\Delta m_2^2&=\frac{1}{32\pi^2}\, G(m_H^2)\left[
4g_Z^2 \cos 2\beta+2g_Z^2 \cos^2\beta+g_2^2\sin^2\beta
\right]\nonumber\\
\Delta m_3^2&=\frac{1}{32\pi^2}\, G(m_H^2)\left[
-2g_Z^2-g_2^2\right]\sin\beta\cos\beta   \,\,.
\label{scalars2}
\end{align}
%


\section{\sc One-loop scale invariance of the matching}
\label{appendixB}

\noindent
In section \ref{toy} we showed that by construction the one-loop matching conditions obtained via one-loop RG-improved effective potentials are independent of the choice of $\mathcal Q_m$. Here we use this property to check our finding \eqref{Deltam2}.

The matching condition~\eqref{deltam}  led to (cf.~section~\ref{2toy})
\be
m^2_{LE}(\mathcal Q_m)=m^2(\mathcal Q_m)  [1+2 \Delta Z_h(\mathcal Q_m)]+ \Delta m^2 \ ,
\label{deltam_gen}
\ee
where [cf.~Eqs.~\eqref{matchingrad} and \eqref{output}]
 \bea
m^2&=&- m_1^2 \cos^2\beta- m_2^2 \sin^2\beta+ m_3^2 \sin 2\beta~, \\
\Delta m^2&=&- \Delta m_1^2 \cos^2\beta- \Delta m_2^2 \sin^2\beta+ \Delta m_3^2 \sin 2\beta~, \\
\tan 2\beta&=&2 \widetilde m_3^2/(\widetilde m_2^2-\widetilde m_1^2)~. \label{tan2beta}
\eea
The one-loop scale invariance imposes that in Eq.~\eqref{deltam_gen}
the one-loop $\mathcal Q_m$ dependence of the right hand side 
is the same of the SM. Here we check this issue.

The total derivative of the right hand side (RHS)~in Eq.~\eqref{deltam_gen} with
respect to $\log\mathcal Q_m$ is given by~\footnote{Note that the contribution to
  $\beta_{RHS}$ coming from derivatives of the angle $\beta(\mathcal Q_m)$ cancels out after imposing Eq.~\eqref{tan2beta}.}
\be
\beta_{RHS}=- \beta_{\widetilde m_1^2} \cos^2\beta-\beta_{\widetilde m_2^2} \sin^2\beta+\beta_{\widetilde m_3^2} \sin 2\beta + m^2 (1+2\gamma_{h}-2\gamma_{h_{LE}})~ ,
\label{rhs}
\ee
with
\be
\widetilde m_I^2=m_I^2+\Delta m_I^2= m_I^2+\sum_r \Delta m_{I,r}^2~,\qquad\beta_{\widetilde  m_I^2}=\beta_{m_I^2}+\beta_{\Delta m_I^2}\equiv \beta_{m_I^2}-\sum_r\Delta_I^r~ ,
\label{total}
\ee
where the index $r$ runs over the heavy fields as in
Eq.~\eqref{threshEFF}. The (one-loop) contribution $\Delta m_I^r$ can
hence be deduced from the one-loop $\mathcal Q$ dependence of $\Delta
m^2_{I,r}$. Each quantity $\Delta m^2_{I,r}$ is provided in
appendix~\ref{appendixA} and can be decomposed into its logarithmic
and non-logarithmic contributions as follows:
\be
\Delta m_{I,r}^2 =\Delta^r_\ell m_{I}^2+\Delta_f^r m_{I}^2\equiv\frac{1}{32\pi^2}\ell_{I,r}+ \frac{1}{32\pi^2}f_{I,r}      \, .
\label{eq-decomp}
\ee
In particular $\ell_{I,r}= 32\pi^2\Delta m_{I,r}^2(G\to \widetilde
G)$, with $\widetilde G(x^2)\equiv x^2 \log (x^2/\mathcal Q^2)$,
represents all terms proportional to logarithms of squared masses over
$\mathcal Q_m^2$, and $f_{I,r}= 32\pi^2\Delta m_{I,r}^2(G\to \overline
G)$ with $\overline G(x^2)\equiv- x^2 (1+\delta) $ contains all
non-logarithmic terms ($\delta=0$ in the $\overline{MS}$ and
$\overline{DR}$).  At one loop each $f_{I,r}$ is thus independent of
$\mathcal Q_m$, and it results $\Delta_I^r=\partial \Delta
m_{I,r}^2/\partial \mathcal Q$.

The values of $\sum_r\Delta_I^r$ are determined in steps
  considering the contributions computed in appendix~\ref{appendixA}
  for each sector. It comes out from sfermions
\begin{align}
 (16\pi^2)  \Delta_1^{\widetilde f}&=6Y_b^2(m_Q^2+m_D^2+A_b^2)+2Y_\tau^2(m_L^2+m_E^2+A_\tau^2)+6Y_t^2\mu^2\nonumber\\
&-\frac{3}{5}g_1^2(m_Q^2-2m_U^2+m_D^2-m_L^2+m_E^2)~,
\nonumber\\
 (16\pi^2)  \Delta_2^{\widetilde f}&=6Y_t^2(m_Q^2+m_U^2+A_t^2)+6Y_b^2\mu^2+2h_\tau^2\mu^2
\nonumber\\
&+\frac{3}{5}g_1^2(m_Q^2-2m_U^2+m_D^2-m_L^2+m_E^2)~,
\nonumber\\
 (16\pi^2)  \Delta_3^{\widetilde f}&=6Y_t^2\mu A_t+6Y_b^2\mu A_b+2 Y_\tau^2\mu A_\tau~,
\end{align}
from charginos
\begin{align}
 (16\pi^2)  \Delta_1^{\widetilde \chi^\pm}&=\Delta_2^{\widetilde \chi^\pm}=-4 g_2^2(M_2^2+\mu^2)~,
\nonumber\\
 (16\pi^2)  \Delta_3^{\widetilde \chi^\pm}&=4g_2^2\mu M_2~ ,
\end{align}
from neutralinos
\begin{align}
 (16\pi^2)  \Delta_1^{\widetilde \chi^0}&=\Delta_2^{\widetilde \chi^0}=-2g_2^2(M_2^2+\mu^2)-\frac{6}{5}g_1^2(M_1^2+\mu^2)~ ,
\nonumber\\
 (16\pi^2)  \Delta_3^{\widetilde \chi^0}&=2g_2^2M_2\mu+\frac{6}{5}g_1^2 M_1\mu ~,
\end{align}
and from the scalar, pseudoscalar and charged Higgs sector
\begin{align}
 (16\pi^2)  \Delta_1^{H}&=\frac{m_1^2+m_2^2+m^2}{m_1^2+m_2^2+2m^2}\left[2g_Z^2(3m_1^2-2m_2^2+m^2)+g_2^2(m_2^2+m^2)  \right]~ ,
\nonumber\\
 (16\pi^2)  \Delta_2^{H}&=\frac{m_1^2+m_2^2+m^2}{m_1^2+m_2^2+2m^2}\left[2g_Z^2(3m_2^2-2m_1^2+m^2)+g_2^2(m_1^2+m^2)  \right]~ ,
\nonumber\\
 (16\pi^2)  \Delta_3^{H}&=-\frac{m_1^2+m_2^2+m^2}{m_1^2+m_2^2+2m^2}\left[2g_Z^2+g_2^2   \right]m_3^2 ~.
\end{align}

Now using Eq.~\eqref{total} and the MSSM one-loop $\beta$-functions~\cite{Castano:1993ri}
\begin{align}
(16\pi^2) \beta_{m_1^2}&=6Y_b^2(m_Q^2+m_D^2+m_1^2+A_b^2)+2Y_\tau^2(m_L^2+m_E^2+m_1^2+A_\tau^2)\nonumber\\
&+(6Y_t^2-6g_2^2-\frac{6}{5}g_1^2)\mu^2-6g_2^2M_2^2-\frac{6}{5}g_1^2M_1^2-\frac{3}{5}g_1^2 S~ ,\nonumber\\
(16\pi^2) \beta_{m_2^2}&=6Y_t^2(m_Q^2+m_U^2+m_2^2+A_t^2)\nonumber\\
&+(6Y_b^2+2Y_\tau^2-6g_2^2-\frac{6}{5}g_1^2)\mu^2-6g_2^2M_2^2-\frac{6}{5}g_1^2M_1^2+\frac{3}{5}g_1^2 S~ ,\nonumber\\
(16\pi^2) \beta_{m_3^2}&=(3Y_t^2+3Y_b^2+Y_\tau^2-3g_2^2-\frac{3}{5}g_1^2)m_3^2\nonumber\\
&+(6Y_t^2A_t^2+6Y_b^2A_b+2Y_\tau^2A_\tau+6g_2^2M_2+\frac{6}{5} g_1^2 M_1)\mu ~,
\end{align}
where $ S=m_Q^2-2m_U^2+m_D^2-m_L^2+m_E^2+m_2^2-m_1^2$, we can
write
\begin{align}
(16\pi^2) \beta_{\widetilde m_1^2}&=(6Y_b^2+2Y_\tau^2-2g_Z^2-g_2^2)m_1^2
-m^2[ (2g_Z^2+g_2^2)\sin^2\beta+6g_Z^2 \cos 2\beta]
\nonumber\\
(16\pi^2) \beta_{\widetilde m_2^2}&=(6Y_t^2-2g_Z^2-g_2^2)m_2^2
-m^2[ (2g_Z^2+g_2^2)\cos^2\beta-6g_Z^2\cos 2\beta]
\nonumber\\
(16\pi^2) \beta_{\widetilde m_3^2}&=(3Y_t^2+3Y_b^2+Y_\tau^2-2g_Z^2-g_2^2)m_3^2
-m^2(2g_Z^2+g_2^2)\sin\beta \cos\beta \ .
\end{align}
Plugging these expressions in Eq.~\eqref{rhs} yields
\begin{align}
\beta_{RHS}&=6Y_t^2(\sin\beta \cos\beta \,m_3^2-\sin^2\beta\, m_2^2)+(6Y_b^2+2Y_\tau^2)(\sin\beta \cos\beta\, m_3^2-\cos^2\beta\, m_1^2)\nonumber\\
&-(2g_Z^2+g_2^2)(-\cos^2\beta m_1^2-\sin^2\beta m_2^2+\sin 2\beta m_3^2)+
(6g_Z^2 c_{2\beta}^2+2\Delta\gamma) m^2\ ,
\end{align}
where
$\Delta\gamma=\gamma_{h}-\gamma_{LE}=\gamma_{1}\cos^2\beta+\gamma_{2}\sin^2\beta-\gamma_{h_{LE}}=-\frac{3}{2}(\frac{1}{5}g_1^2+g_2^2)$.
(This difference between $\gamma_{h}$ and $\gamma_{LE}$ is caused by
the heavy electroweakinos.) Finally one imposes the tree-level
matching conditions of Eqs.~(\ref{matching1}) and \eqref{treeLevel},
and obtains
\be
\beta_{RHS}=\left(6y_t^2+6y_b^2+2y_\tau^2+6\lambda-\frac{9}{10}g_1^2-\frac{9}{2}g_2^2\right)m^2
\label{SMm2}~ .
\ee
Eq.~(\ref{SMm2}) thus agrees with the $\beta$ function of the
quadratic term in the SM~\cite{Luo:2002ey}. Note that in
Eq.~(\ref{SMm2}) the use of tree-level relations within the one-loop
$\beta$ functions is justified by the fact that we are working at
one-loop order.  

Before concluding, notice that the splitting
\eqref{eq-decomp} turns out to be useful also for $\Delta m^2$. In
such a case it reads as
\begin{align}
\label{}
\Delta m^2&=\Delta_\ell m^2+\Delta_f m^2~ ,\nonumber\\
-\Delta_\ell m^2&=\Delta_\ell m_1^2 \cos^2\beta+\Delta_\ell m^2_2 \sin^2\beta-\Delta_\ell m^2_3 \sin 2\beta~,\nonumber\\
-\Delta_f m^2&=\Delta_f m^2_1 \cos^2\beta+\Delta_f m^2_2 \sin^2\beta-
  \Delta_f m^2_3 \sin 2\beta~
\end{align}
although, as we already noticed, the splitting into a logarithmic (or
one-loop scale dependent) and non-logarithmic (or one-loop scale
independent) is not uniquely defined since e.g.~in the function
$[G(x^2)-G(y^2)]/(x^2 -y^2)$ the splitting process does not commute
with the limit $x\to y$ (cf.~Eq.~\eqref{primera}).




\begin{thebibliography}{99}

\bibitem{SM}
  S.~L.~Glashow,
  Nucl.\ Phys.\  {\bf 22} (1961) 579;
  F.~Englert and R.~Brout,
  Phys.\ Rev.\ Lett.\  {\bf 13} (1964) 321;
  P.~W.~Higgs,
  Phys.\ Rev.\ Lett.\  {\bf 13} (1964) 508;
  S.~Weinberg,
  Phys.\ Rev.\ Lett.\  {\bf 19} (1967) 1264;
  A. Salam (1968), N. Svartholm ed., "Elementary Particle Physics: Relativistic Groups and Analyticity", Eighth Nobel Symposium, p. 367.

\bibitem{Gildener:1976ih}
  E.~Gildener and S.~Weinberg,
  Phys.\ Rev.\ D {\bf 13} (1976) 3333;
  E.~Gildener,
  Phys.\ Rev.\ D {\bf 14} (1976) 1667.

\bibitem{Tavares:2013dga}
  G.~Marques Tavares, M.~Schmaltz and W.~Skiba,
  Phys.\ Rev.\ D {\bf 89} (2014) 1,  015009
  [arXiv:1308.0025 [hep-ph]].

\bibitem{Nilles:1983ge}
  H.~P.~Nilles,
  Phys.\ Rept.\  {\bf 110} (1984) 1.

\bibitem{Draper:2013oza}
  P.~Draper, G.~Lee and C.~E.~M.~Wagner,
  Phys.\ Rev.\ D {\bf 89} (2014) 5,  055023
  [arXiv:1312.5743 [hep-ph]];
  E.~Bagnaschi, G.~F.~Giudice, P.~Slavich and A.~Strumia,
  JHEP {\bf 1409} (2014) 092
  [arXiv:1407.4081 [hep-ph]].

\bibitem{Baer:2012cf}
  H.~Baer, V.~Barger, P.~Huang, D.~Mickelson, A.~Mustafayev and X.~Tata,
  Phys.\ Rev.\ D {\bf 87} (2013) 11,  115028
  [arXiv:1212.2655 [hep-ph]].


\bibitem{Sher:1993mf}
  M.~Sher,
  Phys.\ Lett.\ B {\bf 317} (1993) 159
   [Addendum-ibid.\ B {\bf 331} (1994) 448]
  [hep-ph/9307342];
  G.~Altarelli and G.~Isidori,
  Phys.\ Lett.\ B {\bf 337} (1994) 141;
  J.~A.~Casas, J.~R.~Espinosa and M.~Quiros,
  Phys.\ Lett.\ B {\bf 342} (1995) 171
  [hep-ph/9409458];
  J.~R.~Espinosa and M.~Quiros,
  Phys.\ Lett.\ B {\bf 353} (1995) 257
  [hep-ph/9504241];
  J.~A.~Casas, J.~R.~Espinosa and M.~Quiros,
  Phys.\ Lett.\ B {\bf 382} (1996) 374
  [hep-ph/9603227];
%
  G.~Isidori, G.~Ridolfi and A.~Strumia,
  Nucl.\ Phys.\ B {\bf 609} (2001) 387
  [hep-ph/0104016];
  G.~Degrassi, S.~Di Vita, J.~Elias-Miro, J.~R.~Espinosa, G.~F.~Giudice, G.~Isidori and A.~Strumia,
  JHEP {\bf 1208} (2012) 098
  [arXiv:1205.6497 [hep-ph]];
  D.~Buttazzo, G.~Degrassi, P.~P.~Giardino, G.~F.~Giudice, F.~Sala, A.~Salvio and A.~Strumia,
  JHEP {\bf 1312} (2013) 089
  [arXiv:1307.3536 [hep-ph]];
  V.~Branchina and E.~Messina,
  Phys.\ Rev.\ Lett.\  {\bf 111} (2013) 241801
  [arXiv:1307.5193 [hep-ph]];
V.~Branchina, E.~Messina and M.~Sher,
  Phys.\ Rev.\ D {\bf 91} (2015) 013003
  [arXiv:1408.5302 [hep-ph]];
%
  J.~R.~Espinosa, G.~F.~Giudice, E.~Morgante, A.~Riotto, L.~Senatore, A.~Strumia and N.~Tetradis,
  arXiv:1505.04825 [hep-ph].




\bibitem{Veltman:1980mj}
  M.~J.~G.~Veltman,
  Acta Phys.\ Polon.\ B {\bf 12} (1981) 437.

\bibitem{Ford:1992mv}
  C.~Ford, D.~R.~T.~Jones, P.~W.~Stephenson and M.~B.~Einhorn,
  Nucl.\ Phys.\ B {\bf 395} (1993) 17
  [hep-lat/9210033].

\bibitem{Coleman:1973jx}
  S.~R.~Coleman and E.~J.~Weinberg,
  Phys.\ Rev.\ D {\bf 7} (1973) 1888.
  

\bibitem{Bando:1992wy}
  M.~Bando, T.~Kugo, N.~Maekawa and H.~Nakano,
  Prog.\ Theor.\ Phys.\  {\bf 90} (1993) 405
  [hep-ph/9210229];
%
  H.~Nakano and Y.~Yoshida,
  Phys.\ Rev.\ D {\bf 49} (1994) 5393
  [hep-ph/9309215];
%
  C.~Ford,
  hep-th/9609127;
%
  C.~Ford and C.~Wiesendanger,
  Phys.\ Lett.\ B {\bf 398} (1997) 342
  [hep-th/9612193];
%
  J.~A.~Casas, V.~Di Clemente and M.~Quiros,
  Nucl.\ Phys.\ B {\bf 553} (1999) 511
  [hep-ph/9809275];
  J.~A.~Casas, V.~Di Clemente and M.~Quiros,
  Nucl.\ Phys.\ B {\bf 581} (2000) 61
  [hep-ph/0002205].

\bibitem{Feng:1999zg}
  J.~L.~Feng, K.~T.~Matchev and T.~Moroi,
  Phys.\ Rev.\ D {\bf 61} (2000) 075005
  [hep-ph/9909334];
  J.~L.~Feng, K.~T.~Matchev and D.~Sanford,
  Phys.\ Rev.\ D {\bf 85} (2012) 075007
  [arXiv:1112.3021 [hep-ph]].



\bibitem{Delgado:2014vha}
 C.~E.~M.~Wagner,
  Nucl.\ Phys.\ B {\bf 528} (1998) 3
  [hep-ph/9801376];
%
  A.~Delgado, M.~Quiros and C.~Wagner,
  JHEP {\bf 1404} (2014) 093
  [arXiv:1402.1735 [hep-ph]].


\bibitem{Carena:2008rt}
  M.~Carena, G.~Nardini, M.~Quiros and C.~E.~M.~Wagner,
  JHEP {\bf 0810} (2008) 062
  [arXiv:0806.4297 [hep-ph]].

\bibitem{Einhorn:1992um}
  M.~B.~Einhorn and D.~R.~T.~Jones,
  Phys.\ Rev.\ D {\bf 46} (1992) 5206.

\bibitem{Masina:2013wja}
  I.~Masina and M.~Quiros,
  Phys.\ Rev.\ D {\bf 88} (2013) 093003
  [arXiv:1308.1242 [hep-ph]].

\bibitem{Masina:2012tz}
  I.~Masina,
  Phys.\ Rev.\ D {\bf 87} (2013) 053001
  [arXiv:1209.0393 [hep-ph]]. 

\bibitem{AbdusSalam:2011fc}
  S.~S.~AbdusSalam, B.~C.~Allanach, H.~K.~Dreiner, J.~Ellis, U.~Ellwanger, J.~Gunion, S.~Heinemeyer and M.~Kraemer {\it et al.},
  Eur.\ Phys.\ J.\ C {\bf 71} (2011) 1835
  [arXiv:1109.3859 [hep-ph]].

\bibitem{Delgado:2013gza}
  A.~Delgado, M.~Garcia and M.~Quiros,
  Phys.\ Rev.\ D {\bf 90} (2014) 1,  015016
  [arXiv:1312.3235 [hep-ph]].





\bibitem{Delgado:2014kqa}
A.~Delgado, G.~Nardini and M.~Quiros,
  JHEP {\bf 1204} (2012) 137
  [arXiv:1201.5164 [hep-ph]];
%
  A.~Delgado, M.~Quiros and C.~Wagner,
  Phys.\ Rev.\ D {\bf 90} (2014) 3,  035011
  [arXiv:1406.2027 [hep-ph]].


\bibitem{Giudice:1998bp}
  G.~F.~Giudice and R.~Rattazzi,
  Phys.\ Rept.\  {\bf 322} (1999) 419
  [hep-ph/9801271].


\bibitem{Antoniadis:1998sd}
  I.~Antoniadis, S.~Dimopoulos, A.~Pomarol and M.~Quiros,
  Nucl.\ Phys.\ B {\bf 544} (1999) 503
  [hep-ph/9810410];
  A.~Delgado, A.~Pomarol and M.~Quiros,
  Phys.\ Rev.\ D {\bf 60} (1999) 095008
  [hep-ph/9812489];
  R.~Barbieri, L.~J.~Hall and Y.~Nomura,
  Phys.\ Rev.\ D {\bf 63} (2001) 105007
  [hep-ph/0011311];
  A.~Delgado and M.~Quiros,
  Nucl.\ Phys.\ B {\bf 607} (2001) 99
  [hep-ph/0103058];
  A.~Delgado, G.~von Gersdorff and M.~Quiros,
  Nucl.\ Phys.\ B {\bf 613} (2001) 49
  [hep-ph/0107233];
  M.~Quiros,
  hep-ph/0302189;
  S.~Dimopoulos, K.~Howe and J.~March-Russell,
  Phys.\ Rev.\ Lett.\  {\bf 113} (2014) 111802
  [arXiv:1404.7554 [hep-ph]].



\bibitem{ArkaniHamed:2004yi}
G.~F.~Giudice and A.~Romanino,
  Nucl.\ Phys.\ B {\bf 699} (2004) 65
   [Erratum-ibid.\ B {\bf 706} (2005) 65]
  [hep-ph/0406088];
%
  N.~Arkani-Hamed, S.~Dimopoulos, G.~F.~Giudice and A.~Romanino,
  Nucl.\ Phys.\ B {\bf 709} (2005) 3
  [hep-ph/0409232].

\bibitem{Gunion:2002zf}
  J.~F.~Gunion and H.~E.~Haber,
  Phys.\ Rev.\ D {\bf 67} (2003) 075019
  [hep-ph/0207010];
  A.~Delgado, G.~Nardini and M.~Quiros,
  JHEP {\bf 1307} (2013) 054
  [arXiv:1303.0800 [hep-ph]];
  M.~Carena, I.~Low, N.~R.~Shah and C.~E.~M.~Wagner,
  JHEP {\bf 1404} (2014) 015
  [arXiv:1310.2248 [hep-ph]].

\bibitem{Ibrahim:2002zk}
  T.~Ibrahim and P.~Nath,
  Phys.\ Rev.\ D {\bf 66} (2002) 015005
  [hep-ph/0204092].

\bibitem{Castano:1993ri}
  D.~J.~Castano, E.~J.~Piard and P.~Ramond,
  Phys.\ Rev.\ D {\bf 49} (1994) 4882
  [hep-ph/9308335];
  S.~P.~Martin and M.~T.~Vaughn,
  Phys.\ Rev.\ D {\bf 50} (1994) 2282
   [Erratum-ibid.\ D {\bf 78} (2008) 039903]
  [hep-ph/9311340].

\bibitem{Luo:2002ey}
  M.~x.~Luo and Y.~Xiao,
  Phys.\ Rev.\ Lett.\  {\bf 90} (2003) 011601
  [hep-ph/0207271].


\end{thebibliography}
\end{document}